\documentclass[12pt]{article} 
\usepackage{hyperref} 
     \usepackage{amsmath}
      \usepackage{amssymb}
  \usepackage{graphicx}
  \newlength{\abstractwidth}
  \usepackage{mciteplus} 
\usepackage{subcaption}
 \usepackage{bbold}

  \newcommand{\be}{\begin{equation}}
  \newcommand{\bea}{\begin{eqnarray}}
  \newcommand{\eea}{\end{eqnarray}}
  \newcommand{\beq}{\begin{equation}}
  \newcommand{\ee}{\end{equation}}
  \newcommand{\eeq}{\end{equation}}

\def\la{\label}
\def \at {\biggl{\vert}}

\def\32{{3 \over 2 } }

  \def\ba{\begin{eqnarray}}
  \def\ea{\end{eqnarray}}

 \def\simleq{\; \raise0.3ex\hbox{$<$\kern-0.75em
      \raise-1.1ex\hbox{$\sim$}}\; }
 \def\simgeq{\; \raise0.3ex\hbox{$>$\kern-0.75em
      \raise-1.1ex\hbox{$\sim$}}\; }
 
\def\nref#1{(\ref{#1})}

\def \l {\left(}
\def \r {\right)}

\def \lll {\langle}
\def \rrr {\rangle}

\def \wt {\widetilde}

\def \al {\alpha}
\def \bt {\beta}
\def \ga {\gamma}

\def \ep {\epsilon}
\def \th {\theta}

\def \da {\dagger}

\def \pr {\partial}
\def \ra {\rightarrow}
\def \Lc {\mathcal{L}}

\def \al {\alpha}
\def \bt {\beta}
\def \ga {\gamma}

\def \lm {\lambda}

\def \ep {\epsilon}
\def \th {\theta}

\def \Lc {\mathcal{L}}
\def \Rc {\mathcal{R}}

\def \da {\dagger}

\def \pr {\partial}
\def \ra {\rightarrow}

\def \oh {\cfrac{1}{2}}

\def \da {\dagger}

\newcommand\diag{\operatorname{diag}}
\newcommand\Tr{\operatorname{Tr}}

\newcommand{\dist}[2]{(\vec{B}_{#1}-\vec{B}_{#2})}

\begin{document}

\begin{titlepage}
 % \rightline{}
  \bigskip

  \bigskip\bigskip

  \bigskip

\begin{center}
%\centerline
%{\Large \bf {  Pure states in the SYK model \\
%\bigskip
 %and nearly-$AdS_2$ gravity  \\ \bigskip
  % }}
 \bigskip
%\centerline
{\Large \bf { To gauge or not to gauge? }} 
    \bigskip
\bigskip
\end{center}

  \begin{center}

 \bf{  Juan Maldacena$^1$ and Alexey Milekhin$^2$    }
 \\
  \bigskip \rm
\bigskip
\rm
 \bigskip
    $^1$Institute for Advanced Study,  Princeton, NJ 08540, USA  \\
\bigskip
 $^2$Jadwin Hall, Princeton University,  Princeton, NJ 08540, USA \\

 % \bf {Write authors  }
  \bigskip \rm
\bigskip
 
 %   Institute for Advanced Study,  Princeton, NJ 08540, USA  \\
\rm

\bigskip
\bigskip

% \vspace{2cm}
  \end{center}

 \bigskip\bigskip
  \begin{abstract}
%  We study the non-singlet sector of D0 brane quantum mechanics. We conjecture that in the large $N$ limit there remains a finite
%energy gap between singlet and non-singlet excitations. Holographically this happens because all non-singlets live near the boundary.
%The adjoint sector can by analysed by studying the motion of a folded string coming from the boundary, and we find the corresponding spectrum.
% Also we analyse this conjecture from the Hagedorn transition perspective.  

The D0 brane, or BFSS, matrix model is a quantum mechanical theory with an interesting gravity dual. 
We consider a variant of this model where we treat the $SU(N)$ symmetry as a global symmetry,  rather than as a gauge symmetry. 
This  variant  contains new non-singlet states. We consider the impact of these new states on its gravity dual. 
 We argue that  the gravity dual is essentially the same as the one for the original matrix model. The
  non-singlet states have higher energy at strong coupling and are therefore  dynamically suppressed.

 \medskip
  \noindent
  \end{abstract}
\bigskip \bigskip \bigskip

  \end{titlepage}

  %  \starttext \baselineskip=17.63pt \setcounter{footnote}{0}
   \tableofcontents

 % \sc

\section{Introduction} 

Many   examples of the holographic correspondence  involve   very strongly coupled 
large $N$ gauge theories  which are dual to a bulk Einstein gravity theory \cite{Maldacena:1997re,Gubser:1998bc,Witten:1998qj}. In such theories, 
 the gauge symmetry leads to a reduction in the naive number of low dimension  operators from $N^2$ to an order one number. 
The D0 brane matrix model \cite{deWit:1988wri}, also known as BFSS model \cite{Banks:1996vh}, is an example  of such gauge/gravity duality \cite{Itzhaki:1998dd}.
%The matrix model itself is a simple reduction of $d=4$ $\mathcal{N}=4$ super Yang--Mills theory to one dimension. 
In a 0+1 dimensional theory, the only role of the gauge symmetry is to impose an $SU(N)$ singlet constraint. Therefore, we can consider an alternative
 model where we set $A_t=0$. The theory now has a global $SU(N)$ symmetry. If we impose a ``Gauss Law'' constraint restricting to $SU(N)$ singlets, then we recover the gauged model. In this paper we study the properties of the model where we {\it do not} impose this singlet constraint.

 At first sight, one might think that not imposing this constraint leads to many more states, of order $N^2$ of them. The presence of these new states could modify
 the properties of the system substantially. This is indeed correct  in the weakly coupled regime. However, we will argue that in the strongly coupled regime we have essentially the same gravity dual description as for the gauged model. 
 
\begin{figure}[h!]
\centering
\includegraphics{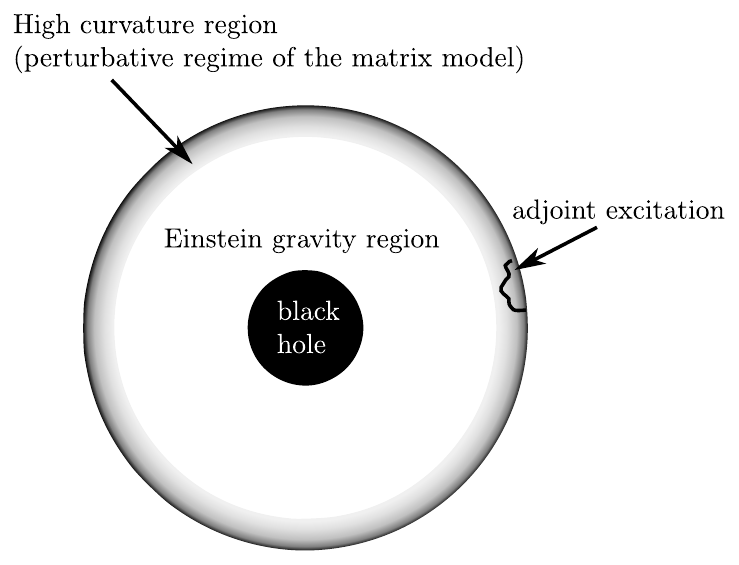}
\caption{Sketch of the gravity solution at finite temperature. The shaded region near the boundary is highly curved. Moving further inwards we find a region of lower curvature that can be described by Einstein gravity. We will argue that 
the lowerst energy  non-singlet excitations live purely  in the shaded region  and have an energy that roughly corresponds to that of a massive string state located in the interface between the two.  }
\label{fig:boundary}
\end{figure}

 In this matrix model, 
 the coupling constant, $g^2$,   has  dimensions  of $(\text{mass})^3$. Therefore it is weakly coupled at high energies and strongly coupled at 
 low energies.  Correspondingly, the gravity dual has a curvature that depends on the radial position. Near the boundary it is highly curved, but away from the boundary we have a low curvature region where we can trust Einstein gravity. See Figure \ref{fig:boundary}. 
 This low curvature  region corresponds to the energy scales where the matrix model is strongly coupled.  
 
  We will argue/conjecture  that the gravity picture of the non-singlet states is the following. The  non-singlet states have an energy of order 
 the order $\lambda^{1/3} = (g^2 N)^{1/3}$ and are located in the high curvature region, away from the region that is described by  Einstein gravity, 
 see Figure \ref{fig:boundary}. 
% More precisely, the full gravity solution has a curvature that varies with radial position. It becomes highly curved near the boundary, so that the region near the boundary cannot be described in gravity. This region can be described by perturbation theory in the matrix model. It turns out that non-singlets are confined to live in this region and have a hard time penetrating into the region described by Einstein gravity. See figure XXX [	REFER TO FIGURE]	
 In the planar approximation, we also have states corresponding to additional excitations of these non-singlet states 
 which can be represented as folded strings with their ends stuck to the highly curved region near the boundary.  
 At finite temperature we can further have non-singlet states that correspond to black holes with strings that come in from the boundary and end on the black hole, see Figure \ref{folded_string} (b,c). 

 We have not derived this picture rigorously, we will simply present some plausibility arguments and consistency checks. 
  %  The authors of
  In  \cite{Numeric}, 
 Berkowitz, Hanada, Rinaldi and Vranas present more   evidence  supporting this picture by performing a direct numerical simulation of the ungauged model\footnote{ We thank the authors of \cite{Numeric} for detailed explanations on their computations and for ongoing discussions. }.
 
 In \cite{Gross:1990md}, a similar conclusion was reached  for the ``double scaling''  limit of a single matrix quantum mechanics for low enough temperatures
 See also \cite{Marchesini:1979yq,Boulatov:1991xz,Kazakov:2000pm,Maldacena:2005hi,Fidkowski:2005ck} for further discussion of non-singlets in that model. 

We were  motivated to ask the   question in the title by  the Gurau-Witten tensor models \cite{Gurau:2011xp,Witten:2016iux}. There
  one has the choice of either imposing or not imposing a singlet 
constraint. It is sometimes thought that the models with a singlet constraint would be  more likely to have a gravity or string dual.
 Our main message is that the existence of a local gravity dual is independent of whether we do or do not impose this constraint.
  Einstein-like gravity is  associated to very strong  interactions but not to the presence or absence of the boundary theory gauge symmetry (or gauge redundancies).

%Near the boundary of this geometry $S^8$ becomes very small and thus we can not trust Einstein gravity beyond certain point. 
%Fortunately, this "complicated" region is in fact 
%the UV region of the matrix model where it is weakly coupled. We will call the region where both gravity and perturbation theory break down the correspondence region.
%Gravity predictions with the perturbative calculations have to agree in this region. This will allow us to estimate the gap between singlets and non-singlets.

%We want to explore the relevance of the gauge symmetry to the holographic duality. 
%In quantum mechanics gauge symmetry is a global symmetry plus a singlet constraint. What if we do not impose the singlet constraint? 
%How different is the physics? What can we say about the gravity duals of the non-singlets?
%Our main message, is that simple non-singlets such as adjoints and other states with relatively small representations have the same 
%gravity duals except for extra excitations that live near the boundary. 

When we consider the ungauged model we break supersymmetry, since in the original model the algebra only closes up to gauge transformations. 
Nevertheless the modified algebra can be used to argue that the energy is positive, even for non-singlet states. Of course, singlet states are the same as those 
of the gauged model. For non-singlet states,  the lowest energy state  appears to be when all branes are separated by a large amount (namely,  the matrices get large 
diagonal expectation values). In this regime,   the 
non-zero $SU(N)$  charges lead to a kind of angular potential going like $1/X^2$ for diagonal matrices of typical magnitude $X$.
 This potential leads to even larger expectation values for the matrices. 
Nevertheless, for finite temperatures, we expect to have a metastable state where  the expectation values of the matrices are relatively small (or the branes are together), since this state has more entropy. 
This state can be viewed as a  black hole. Our previous remarks on the equality of the gravity configurations applied for these metastable black hole configurations.  
 
We are arguing that non-singlets are energetically disfavored at low energies. This seems to contradict the picture proposed in \cite{Witten1998,Aharony2003} for the 
deconfinement/black hole transition that is based on the idea that the Polyakov loop gets an expectation value. If the only states contributing were singlets we would get no potential for the eigenvalues of the Polyakov loop. 
We will discuss how the two pictures are consistent.  We are lead to a picture where the values of the holonomy indeed break the center symmetry 
 but only by a very small amplitude ``wave'' in the eigenvalue distribution.

This paper is organized as follows. In Section \ref{sec:mmodel} we review the BFSS matrix model and its holograhic dual. Also we describe its massive deformation, 
the BMN or plane wave model. 
% \cite{bmn}. 
In Section \ref{sec:ungauged} we describe how one can remove the singlet constaint and obtain the deformation of the supersymmetry algebra.
We also relate it to the insertion of Wilson loops in the gauged model. 
Section \ref{sec:nonsinglets} is devoted to non-singlets. We first look at the lowest energy excitations of the thermal blackground. We then consider the region where all branes are far away. We also use perturbation theory to find the shifts to the spectrum in the weakly coupled region. 
We also  discuss the thermodynamic properties of the ungauged model. 
We discuss further aspects of the Polyakov loop and thermal phase transitions in  section \ref{sec:hagedorn}. After making some further comments 
we present some conclusions. 
 
\section{The D0 brane matrix model } 
\label{sec:mmodel}
 
In this section we review the D0 brane matrix model and its gravity dual.
 Readers familiar with this material can jump directly to the next section. 

\subsection{The matrix model} 
\label{sec:bfss}
The D0 matrix model \cite{deWit:1988wri}, or BFSS matrix model \cite{Banks:1996vh}, has the action 
\be 
\la{BFSSla}
S =\frac{1}{g^2}  \int dt \Tr \l \oh  (D_tX^I)^2 + \oh \psi_\al D_t \psi_\al + \frac{1}{4} [X^I,X^J]^2 + \oh 
i \psi_\al \ga_{\al \bt}^I [\psi_\bt,X^I] \r
\ee
where all indices are summed over. $I,J=1,\cdots,9$, $\alpha,\bt =1, \cdots 16 $. 
where $\gamma^I$ are nine dimensional gamma matrices which are real, symmetric and traceless\footnote{
We can view them as coming from the ten dimensional Majorana Weyl representation $\gamma^I = \Gamma^0 \Gamma^I$.}. 
 $\psi_\alpha$ are hermitian $N \times N$ matrices, which can be expanded as 
$\psi_\alpha = \psi^r_{\alpha} T^r$ where the $T^r$ are a complete set of hermitian $N\times N$ matrices, and we can think of the $r$ index as a real 
index of the adjoint representation of $U(N)$. 
Then $\psi^r_\alpha$ are Majorarana fermions. We have $16 \times N^2$ Majorana fermions. 
This model is invariant under 16 supersymmetries and also under an $SO(9)$  R-symmetry. 

The model has a $U(N)$ gauge symmety and the derivative is defined as $D_t B = \partial_t B + i[A_t , B] $ where $A_t$ is the gauge field. 
We could choose the gauge where $A_t=0$ and then we have to impose Gauss's law:
\be
\label{Gauss}
G = \frac{i}{2g^2} \l  2 [D_tX^I,X^I] + [\psi_\al,\psi_\al] \r = 0
\ee
It restricts all states to be singlets under the $U(N)$ symmetry. 

Classical zero energy configurations correspond to simultaneously diagonal matrices $X^I$. 
Quantum mechanically, the model has a zero energy bound state. At finite temperature it is expected (from the gravity picture) to have a metastable 
bound state. 
% where the $X^I$ are not simultaneously diagonal. 

It is interesting to ask what the typical size of the matrices $X^I$ is in the ground state or in a thermal state. 
  This was estimated \cite{Polchinski:1999br}  by setting a lower bound for 
   ${ 1 \over N } \Tr[ X^2] $, using virial theorem ideas. That lead to 
  \be \la{TypEig}
   \sqrt{ \langle { 1 \over N } \Tr[ X^2]  \rangle } \sim \lm^{1/3} ~,~~~~~~~ \lambda \equiv g^2  N 
   \ee
 In a heuristic way,   this can also be obtained by dimensional analysis and large $N$ counting if one assumes that $\lambda$ is the only relevant scale (and not the temperature). This result will be particularly useful when we analyze  the gravity solution. 
 % We would like to
%emphasize that the above estimate (\ref{TypEig})
% is very important and has a lot of physical consequencies. We will refer to it several time throughout the paper.

There is a variant of this model where we add mass terms that break $SO(9) \to SO(3) \times SO(6)$ \cite{bmn}. The additional terms in the action are 
\begin{eqnarray}
%\begin{split}
\la{BMNmu}
S_{BMN} &=& S_{ [{\rm from } ~ \nref{BFSSla} ] } + S_\mu ~,
\\
S_\mu &=&  - \frac{1}{g^2} \int dt \Tr \Bigg( \oh \l{\mu \over 3}\r^2 \sum_{a=1 }^3 \l X^a \r^2 + \oh \l {\mu \over 6} \r^2 \sum_{i =4  }^9  (X^i)^2 +
\frac{\mu}{8} \psi \gamma_{123} \psi + 
\cr
 &&~~~~~~~~~~~~~ +i \frac{\mu}{3} \sum_{a,b,c=1}^3  X^a X^b X^c  \ep_{abc} \Bigg)
%\end{split}
\end{eqnarray}
% Once we add \nref{BFSSla} and \nref{BMNmu} we obtain the BMN matrix model\cite{BMN} 
It  also preserves 16 supercharges but with a different supersymmetry
algebra, $SU(2|4)$. 
We can view \nref{BMNmu} as a collection of harmonic oscillators and Majorana fermions with some particular interactions. 
 
 The mass terms remove the  flat directions in the potential. 
Apart from the simplest vacuum with $X^I=0$,  the BMN model also has additional vacua \cite{Dasgupta:2002hx,Maldacena:2002rb,kim}, 
characterized by non-zero $X^a,\ a=1,2,3$ such that:
\beq
\label{fuzzy_vacuum}
i \ep_{abc} X^b X^c = \frac{\mu}{3} X^a
\eeq
This equation is solved by $X^a = \frac{\mu}{3} J^a$, where $J^a$ are $SU(2)$ algebra generators in an $N$-dimensional representation, 
% \footnote{With our normalization we have$\Tr \l J^a J^b \r = \oh \delta^{ab}$ for fundamental representation},
 not neccessarily irreducible. Such solutions
represent a collection of fuzzy spheres.
Although this vaccum breaks $SU(N)$ symmetry, there are no physical Goldstone bosons because of the gauge symmetry.
 We will return to $SU(N)$ Goldstone bosons later in Section \ref{sec:ungauged}
when we discuss the ungauged model.

\subsection{The gravity dual } 

We will be mostly discussing the gravity dual at finite temperature. The  geometry is a solution of ten dimensional type IIA supergavity closely related 
to the near horizon geometry of a charged black hole in ten dimensions \cite{Horowitz:1991cd}. It is given by \cite{Itzhaki:1998dd}
\bea
\label{geometry}
 \frac{ds^2}{\al'} &=&  - \frac{f_0(r) r^{7/2}}{\sqrt{\lm d_0}}  dt^2 +   \sqrt { \lambda d_0 \over r^3 }  \l \cfrac{1}{f_0(r) r^2}  dr^2 + d \Omega^2_8 \r
 \cr
 e^{\phi} &=&  { ( 2 \pi )^2 \over d_0 } { 1 \over N }  \l \frac{\lm d_0}{r^3} \r^{7/4} 
 \cr
 % A_t &=& -\frac{1}{2} \l \frac{r^7}{\lm d_0} - 1 \r
 \tilde A_t &=& { N \over (2\pi)^2 }       \frac{r^7}{\lm^2  d_0}   
 \cr
  f_0(r) &=& 1-\frac{r_0^7}{r^7},  ~~~~~~ \ d_0\equiv 240 \pi^{5},\ ~~~~~ \lm \equiv    g^2 N ,\  % g_{YM}^2 = \frac{g_s}{4 \pi^2 \al'^{3/2}}
 \eea 
%This geometry is the near horizon geometry of a ten dimensional charged black hole.  
where 
$r_0$ and the inverse temperature $\beta=1/T$ are related by\footnote{ We can think of the relation between $\beta$ and $r_0$ as a way to translate between time scales in the matrix model ($\beta$) and radial position in the bulk ($r_0$) \cite{Peet:1998wn}. } 
\be
\label{eq:r0} 
{ 1 \over T} = \beta = \frac{4}{7} \pi \sqrt{\lm d_0} r_0^{-5/2}
\ee
This  geometry has an effective radius of curvature given by the radius of $S^8$
\be
  { R^2_{eff}  \over \alpha'}  =   \sqrt { \lambda d_0 \over r^3 }    % \frac{\sqrt{\lm d_0}}{r^{3/2}}
\ee
which 
is a function of the radial direction. For this reason we can trust \nref{geometry} only in some region of 
the geometry, namely $r \lesssim \lm^{1/3}$. Note that $r$ has units of energy. 
At larger values of $r$, when $  \lm^{1/3} \lesssim r $,  the curvature is high and we cannot trust the gravity solution.
The large $r$ region is where the boundary is and  it corresponds to the UV of the boundary theory.
 In this region the 
 %   However for very large $r \gg \lm^{1/3}$ the 
 matrix model is weakly 
coupled and we can trust   perturbation theory.

The geometry at the horizon of the black hole will be weakly curved as long as 
\be
1 \ll { \lambda \beta^3 }
\ee
There is an additional $N$ dependent
 constraint $\lm \beta^3 \ll N^{10/7}$ on the validity of this IIA supergravity 
 solution that arises when we also demand that the dilaton is not too large at the horizon. 
In this paper, we will imagine that we are in the `t Hooft limit where $N$ is taken to be very large compared to other quantities, such as $\lambda$ or $\beta$, or
more precisely $\lambda \beta^3$. So we do not have to worry about this second constraint.

Using the Bekenstein--Hawking formula one can easily find the entropy and free energy\footnote{ The temperature dependence can be recovered from the properties of   \nref{geometry} under rescalings. Namely sending $t \to \eta t$ and $r\to \eta^{-2/5} r $ the metric gets rescaled by an overall factor and the action 
by $S \to \eta^{-9/5}S $, which is also the scaling of the entropy. See Appendix \ref{app:scaling}.} 
\beq
\label{BHentropy}
S = N^2 4^{13/5} 15^{2/5} (\pi/7)^{14/5} \l \frac{T}{\lambda^{1/3}} \r^{9/5} \approx 11.5 N^2 \l \frac{T}{\lambda^{1/3}} \r^{9/5}
\eeq
%The dependence $\sim T^{9/5}$ can easily be recovered using the hyperscaling of the metric (\ref{geometry}). In the 
%Appendix \ref{app:scaling} we will discuss this symmetry in more details. 
%
%
%Below we will also need the expression for the free energy:
\beq
F = N^2 \lambda^{1/3} \frac{5}{14} 4^{13/5} 15^{2/5} (\pi/7)^{14/5} \l \frac{T}{\lambda^{1/3}} \r^{14/5} \approx
7.4 N^2 \lambda^{1/3}  \l \frac{T}{\lambda^{1/3}} \r^{14/5} 
\eeq
These predictions were checked in an increasingly sophisticated set of numerical computations 
\cite{Anagnostopoulos:2007fw,Kabat:2000zv,Hanada:2008ez,Kadoh:2015mka,Catterall:2008yz,Filev:2015hia} culminating in \cite{Berkowitz:2016jlq}, where
also the leading $\alpha'$ corrections were computed\footnote{It is an interesting challenge to match the first correction by computing the full 
tree level $\alpha'^3$ corrections to the tree level  IIA supergravity in the effective action.}.

The gravity dual for the BMN case is a bit more complicated, it has some gapped states described in \cite{Lin:2005nh} and a black hole 
thermal state which looks 
like a deformation of \nref{geometry} \cite{Costa:2014wya}. The magnitude of the deformation involves  $\mu/T$ and it is very small if 
$\mu/T$ is small.  

\subsection{The size of  the matrix versus the size of the Einstein gravity region} 
 \la{PolCal}
     
    It is interesting to translate \nref{TypEig} to the gravity side. On the gravity side we can consider D0 brane probes that sit at particular values of $r$. 
    A string stretching from this brane probe to $r=0$,  or the horizon, has an energy of the order of $r$. 
    Now, if we consider the mass of an off-diagonal mode of a matrix in the diagonal background $m \sim X_{diag}$ we expect to get the same
    energy. It means that the radial direction is related to matrix elements as $r \sim X$.
    Using this procedure to translate between radial positions and matrix eigenvalues, we now ask: What value of $r$ would the scale \nref{TypEig} correspond to?
    Interestingly, it corresponds to a scale $r \sim \lambda^{1/3} $, which is the scale at which the supergravity solution breaks down!
    This important point was emphasized in \cite{Polchinski:1999br}, and we are repeating it because 
    we think it is not widely appreciated. 
     In fact, some  papers in the literature seem to suggest that the typical size of the matrices in the thermal state would be 
     $X \sim r_0$. Note that $r_0 \ll \lambda^{1/3} $ in the region where we can trust gravity. 
    
     This means that the whole Einstein gravity region of Figure \ref{fig:boundary} corresponds to a 
    highly quantum region of the wavefunction for the matrix model. The matrices have large fluctuations. 
    However, these fluctuations are highly correlated. Indeed, via supersymmetric localization, \cite{Asano:2014vba,Asano:2017xiy} computed 
      $ { 1 \over N } \langle \Tr[ ( X^1 + i X^9 )^{2k} ] \rangle$. They found   a much smaller
    answer  agreeing  with naive bulk expectations. Due to the $i$, in this expectation values there are interesting cancellations.

\section{ The ungauged model } 
\label{sec:ungauged}
In this paper we will consider the situation where we set $A_t=0$ and we {\it do not }  impose the $SU(N)$-singlet constraint\footnote{
We could also say that we have a $U(N)$ gauge symmetry. However, since there are no fields charged under the overall $U(1)$,  
 it does not matter whether we gauge or do not gauge the 
overall $U(1)$.}. 
This amounts to treating the $SU(N)$ symmetry as a global symmetry rather than as a  gauge symmetry. 
In higher dimensions, gauging a symmetry introduces extra degrees of freedom. In quantum mechanics it does not. The theory with 
$A_t=0$ is a perfectly well defined theory, with global $SU(N)$ symmetry,  and we can consider it in its own right. This theory has a singlet subsector 
where it identical to the usual one in Section \ref{sec:mmodel}, but it also has non-singlet states whose interpretation in the gravity dual we want to ellucidate. 

It is sometimes said that gauging the $SU(N)$ symmetry reduces the number of operators drastically and that this is important for 
the gravity solution to work. We will see that the gravity solution can be valid whether we gauge the $SU(N)$ symmetry or not. 

\subsection{Lack of supersymmetry} 

Let us define the hamiltonian of the ungauged model to be simply the one obtained from eq. (\ref{BFSSla}) by setting $A_t=0$. 
We can then wonder whether the resulting theory is supersymmetric. We certainly continue to have the operators $Q_a$ that were generating 
the SUSY transformations before:
\be
\label{qep}
Q \ep = -\frac{1}{g^2} \Tr \l  \dot X^I \psi \ga^I \ep +  i \oh [X^K,X^L] \psi \ga^{KL} \ep \r
\ee
where $\ga^{KL}=\oh \l\ga^K \ga^L - \ga^L \ga^K\r$.
We can now ask whether these operators commute with the Hamiltonian. We find 
\be
\label{qh}
[Q_\al, H] = -  \Tr \l \psi_\al G \r 
\ee
We see that the right hand side can be written in terms of the $SU(N)$ symmetry generators, $G$ in \nref{Gauss}. This means that, while \nref{qh}
 vanishes when it acts on  singlet
states, it will be non-vanishing acting on non-singlet states. Therefore we expect that non-singlets will not come in supersymmetry multiplets.
We can also compute the anticommutators 
\be
\label{qq}
\{Q_\al,Q_\bt\} = 2H \delta_{\al \bt} +2 \Tr \l G X^I \r \ga^I_{\al \bt}
\ee
We see that we get   non-zero answers in the right hand side because the supersymmetry transformations only close up to $SU(N)$ transformations. In the
gauged model these are gauge transformations. But in the ungauged model we get a non-zero right hand side. 
Nonetheless, we can still infer some information from this algebra.  

Let us note first, that even for non-siglet states
the energy is non-negative. Indeed, since the supercharges are self-adjoint $Q_\al^\dagger = Q_\al$ and gamma matrices are traceless, 
summing over the spinorial indices leads to
\beq \la{posEn}
32 H = \sum_{\al =1}^{16}  \{Q_\al,Q_\al\} = \sum_{\al =1}^{16}  \{Q^\dagger_\al,Q_\al\} \ge 0
\eeq

\subsection{Supersymmetric version of the ungauged model}

In principle, we could modify the definition of the supercharges so as to have a supersymmetric theory. We do not think that is possible. 
Nevertheless, if we are willing to also redefine the Hamiltonian, then it is possible to preserve some of the supersymmetry. This can be achieved by adding a new
term to the Hamiltonian: 
\be \la{SusyHam}
H_\text{susy} = H - \Tr \l X^1 G \r
\ee
%The supercharges remain the same. 
This breaks the $SO(9)$ symmetry to $SO(8)$, and it preserves half of the supersymmetry, those whose spinorial parameter obeys
\be
%  (\gamma^9+\text{id}) \ep = 0
  (\gamma^1+ {\bf 1 } ) \ep = 0
\label{save}
\ee 
Moreover, now we have the standard supersymmetry algebra:
\be
\label{qq_std}
\{Q \cdot \epsilon , Q \cdot \epsilon'  \} = 2H_\text{susy} \epsilon \cdot \epsilon'
\ee
This might seem surprising at first sight, but there is a simple explanation for the existence of this Hamiltonian. 
In this paper we will concentrate on the model with the original Hamiltonian.  

\subsection{Relation to Wilson loop insertions } 
\label{sec:Wilson}
There is a physical situation that arises in the gauged model  which is very  closely connected to the ungauged model. We can have the original gauged theory and add an external quark in some 
representation $\bar R$, by coupling it through a Wilson line operator in representation $\bar R$. This is very closely related to  the ungauged theory  restricted to  
 the representation $R$ \footnote{ We can only consider representations  transform trivially under the $Z_N$ center of $SU(N)$,
  which are the ones we can get from products of adjoints. }. The only difference is that in the thermal partition function, restricted to representation $R$, 
  we would include a factor of the dimension of
the representation in the ungauged case but not in the gauged case with a Wilson loop. 

The simplest  Wilson loop operators we can consider are $\Tr_{\bar {\text{ R} }} P e^{ i\int A_t dt}$. These  break supersymmetry. 
Another commonly considered operator preserves half of the supersymmetries and has the form $\Tr_{ \bar{\text{ R}} }P e^{ i \int dt ( A_t   +  X^1) } $, where we have picked one of the
scalar fields
\cite{Rey:1998ik,Maldacena:1998im}. The extra term corresponds to the extra term in the Hamiltonian \nref{SusyHam}.
When we add the supersymmetric  Wilson loop in the adjoint representation, in the gravity dual we get a string coming in from the boundary at $X^1 = \infty$ and a string 
going to $X^1 = -\infty$. Equivalently, we can say we have a string anti-string pair with the string pinned on the north pole of the $S^8$ and the anti-string
on the south pole of $S^8$. See Figure \ref{folded_string} (a).
 
 In conclusion, we can translate many of the statements in this paper to statements about insertions of Wilson lines for the original, gauged, model.

\section{Gravity duals of non-singlets} 
\label{sec:nonsinglets}

\begin{figure}[h!]
\centering
\includegraphics{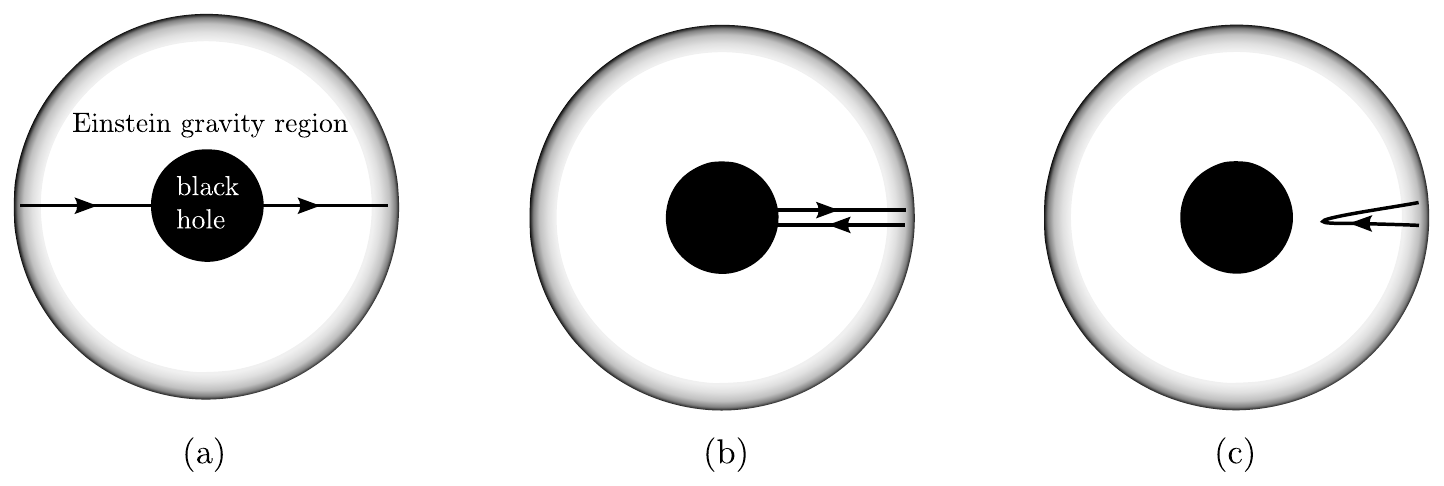}
\caption{(a): The string configuration corresponding to a supersymmetric Wilson line in the adjoint representation.
 (b): The string
and anti-string configuration representing a non-supersymmetric adjoint. We get it from (a) by moving the left string segment to the right side of the figure . (c): We further lower the energy configuration of (b) by 
  moving the tip away from the horizon. The idea is that the end point of this motion is a configuration as in Figure \ref{fig:boundary}.}
\label{folded_string}
\end{figure}

Let us consider first the gravity dual of the adjoint states, states in the adjoint representation of $SU(N)$. 
They  are  described by the gravity dual of the non-supersymmetric Wilson loop $e^{i \int A_t dt }$. 
As pointed out in \cite{Alday:2007he} (see also \cite{Polchinski:2011im}), 
the gravity dual of these Wilson loops differs from the supersymmetric Wilson loops only through the fact that the strings are not
pinned at a particular point on the sphere, but they can move to any point on the sphere. See Figure \ref{folded_string}(b,c). 
In other words, on the boundary of the bulk they obey Neumann, rather than Dirichlet,  boundary conditions in the sphere directions. 
  If we have an adjoint, this means that the string and the anti-string could lower their energy by coming closer together on the sphere. 
If they coincide on the sphere, then we have a folded string whose energy can be lowered by moving the tip further and further to the boundary, see  Figure \ref{folded_string}(c).

According to the gravity solution, the energy of folded  streteched string (at $\beta =\infty$), with its tip momentarily at rest,  is given by 
\be \la{GravEs}
E = {  2 \over 2\pi } \int_{r_{tip}}^{r_\infty}    dr = { 1 \over \pi } \l r_\infty - r_{tip} \r
\ee
This computation, valid in the gravity regime, would suggest that we can lower the energy to zero by moving $r_{tip } \to r_{\infty}$, where $r_{\infty}$ is some
large $r$ cutoff. However, at very large $r$ we cannot trust the gravity computation.  In other words, 
the fact that as $r_{tip} \to r_{\infty}$ the energy goes to zero cannot be trusted when both of these quantities are in the highly curved region. 
Therefore, it could be that even the lowest energy configuration has a non-zero energy. 
What would be a natural value for this energy?
One natural possibility would be to think that $r_\infty \propto \lambda^{1/3}$ which is the value of $r$
where the curvature becomes of the order of the string scale. Furthermore, we can also assume that naive cancellation between $r_{tip}$ and $r_\infty$ does not happen and that we get an energy that is the typical energy of a massive string state at the location given by $r \sim \lambda^{1/3}$. From 
\nref{geometry} we find that this is an energy of the order of $\lambda^{1/3}$. 
 We get the same answer if we use
 dimensional analysis and assume that it will be of the order of the `t Hooft coupling. In both cases we get
\be
\la{Estimate} 
E_\text{min} = C \lambda^{1/3} 
\ee
where $C$ is an unknown numerical constant.  In the next subsection we will present an argument saying that $C>0$.  Note that $C$ cannot be negative because we have argued near \nref{posEn} that the energy should be positive. 
The fact that $C $ is positive is also suggested by the numerical computation in \cite{Numeric}.
 
% At this point this seems somewhat arbitrary.  We will try to offer other pieces of evidence in what follows. 

We can speculate about the temperature corrections to the estimate (\ref{Estimate}).
We expect these to come from the fact that at finite temperature the metric at the transition region, at
 $r\sim \lambda^{1/3}$, will be changed due to the $r_0$ dependent terms in \nref{geometry}. 
 We expect this to produce an extra factor of $(1 + a_1 { r_0^7 \over r^7} )$, where $r\sim \lambda^{1/3}$. Using \nref{eq:r0} we find  then that 
%  By briefly examining the metric
% and the dilaton
%(\ref{geometry}) one can easily see that the temperature enters only as $r_0^7 \sim T^{14/5}$. We expect that 
%string energy will receive corrections of the same order, namely:
\beq
E = C \lambda^{1/3} \l 1 +\tilde a_1  \l  \frac{T^{3 }}{\lambda }  \r^{ 14 \over 15 } +\cdots\r
%  C_1 \frac{T^{14/5}}{\lambda^{14/15}}+C_2 \frac{T^{28/5}}{\lambda^{28/15}} +\dots \r
\eeq
where $\tilde a_1$ is an unknown numerical constant. The main point is that it is small for $T \ll \lambda^{1/3}$. 
% Unfortunarely, it is not clear how to determine the sign of $C_1,C_2,\dots$.

\subsection{Exploring the large $X$ region } 
\la{LaXreg}

In the above discussion we have assumed that the model starts in a state with $X\sim 0$ and then we add the adjoint as a perturbation. 
This is particularly reasonable if the branes are trapped near the origin by thermal effects. 

\begin{figure}[h!]
\centering
\includegraphics[scale=1.5]{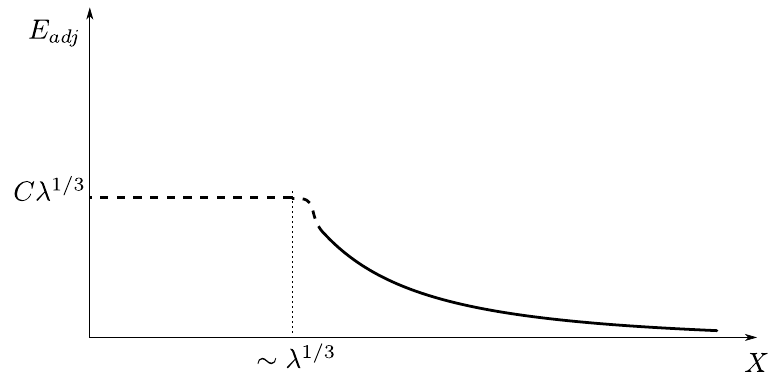}
\caption{Energy of an adjoint state where we explore the flat directions of the potential in \nref{BFSSla}, which correspond to mutually diagonal matrices $X^I$.  The solid line 
denotes the potential computed in the region where we can trust perturbation theory in the matrix model side, which is   $ \lambda^{1/3} \lesssim |X|$. The details depend on the particular form of the diagonal matrices $X$. In this region, the energy comes from the angular momentum along the $SU(N)$ directions 
in the moduli space of vacua of the ungauged model. The horizontal doted line corresponds to the energy of a massive string state in the transition region, as in Figure \nref{fig:boundary}. We expect a smooth transition region in between.   We have not included here the effects of the thermal potential which produces a large dip near $X \sim 0$, or $r\sim r_0$, (for  $T\ll \lambda^{1/3}$) because it  is common to the gauged and ungauged models. 
  }
\label{potential}
\end{figure}

On the other hand, we can set the temperature to zero and consider a situation where all branes are separated from each other. In this case, we can ask 
about the energy of the adjoint state. First we should note that if we do not gauge the symmetry, then we have a manifold of Goldstone modes coming from 
applying the $SU(N)$ transformations to the original configurations. This manifold has an $SU(N)$ symmetry and we can consider a wavefunction 
which is in the adjoint representation under this global $SU(N)$ symmetry. We can think of this as a   configuration which has an $SU(N)$ 
``angular momentum'' along this manifold. The typical radius of this manifold is given by the distance between the branes, call it $X$. Then 
we get an energy which goes like 
\be \la{PotAp}
V \sim  +{  \lm \over X^2} 
\ee  
We discuss and  derive this in more detail in Appendix \ref{sec:bfss_pert}. 
One can view this final formula as analogous to the angular momentum potentials we get when a particle moves in three dimensions in a spherically symmetric potential and with some angular momentum. 
It makes sense to   first freeze  $X$  and then  calculate the   potential \nref{PotAp} for the following reason. 
The effective mass of the $X$ variables is of order $1/g^2 \sim { N \over \lambda }$,  which is large in the `t Hooft limit. Therefore,
 the motion in the $X$ directions produced by \nref{PotAp}  will be relatively slow. This is like the Born-Oppenheimer approximation. 
We can trust \nref{PotAp}
 when $|X|$ is large enough that we can use  perturbation theory in the matrix model. This means that 
$ \lm/X^3 \ll 1$.  If we extend this to the boundary of its regime of validity, namely to $X^3 \sim \lm$, then we find that 
the energy becomes  $\sim \lambda^{1/3}$, in agreement with the estimate \nref{Estimate}.  Figure \ref{potential} shows the form of the  potential when we separate the branes and we are in an adjoint state. The reason we
get a constant when $|X| \lesssim \lambda^{1/3}$ is the picture we suggested in Figure 
\ref{fig:boundary} where the adjoint is localized in the transition region. When the branes are located within the Einstein gravity region they have shed their adjoint
charge, leaving it  as a  string with endpoints in the high curvature region. 
%\footnote{Here we are identifying the degrees of freedom that parametrize the vacuum moduli space in the gravity region, which are separated D0 branes, with the ones that describe it in the perturbative region of the matrix model which are diagonal matrices. EXPLAIN MORE OR MOVE IT TO SOME OTHER LOCATION. }, they have effectively shed their adjoint quantum number, which are simply carried by a massive string state in the transition region, $r\sim \lambda^{1/3}$. 

Note that this transition happens at a value of $r$ that coincides with the 
 size of the ground state wave function (\ref{TypEig}). This also suggests that when the $X$
  have an expectation value  of this size there will be other degrees of freedom that can carry the adjoint quantum numbers. 
  % Here we are imagining that 
 %In fact the energy $\lambda^{1/3}$ can be viewed as the energy of a typical off diagonal mode for a matrix whose diagonal components are of order $\lambda^{1/3}$.

Note that the presence of the potential in Figure \nref{potential} suggests that the adjoint state with 
$X\sim 0$ is unstable and the system is driven to $X \sim \infty$. We think that this is the ultimate fate of adjoint states. On the other hand, at finite temperature the gravity solution shows that 
 thermal effects will trap the branes at 
$X\sim 0$, leading to a metastable minimum.  As we will recall near \nref{eq:emission}, this metastable state is very long lived in the `t Hooft limit, so that we only need to worry about this decay mode at very low temperatures. 

\subsection{Adjoint energies at weak coupling in the BMN matrix model} 

The BFSS matrix model is always strongly coupled at low energies. On the other hand, the BMN matrix model has another scale, given by the mass $\mu$. If we take $\lambda \ll \mu^3$,  we can trust perturbative computations even around  the simplest, $X=0$,  vacuum. 
The 
%It is also possible to find the perturbative energies of non-singlets in the BMN model. In this case the
 expansion parameter is $\lambda/\mu^3$. In the simplest vacuum
we have a collection of bosonic and fermionic harmonic oscillators. The lightest sector with matrix creation--annihilation operators $a^\dagger_i,a_i, \ i=1,\dots,6$ corresponds to $SO(6)$
operators $X^i=i,\ i=1,\dots,6$,  see Appendix \ref{app:bmnper} for details. Each oscillator $a^\dagger_i$ has energy $\mu/6$.
In the gauged $ SU(N) $ model we cannot act with a single creation operator because it would be in the adjoint of $SU(N)$. The first singlet appears for a pair of operators 
$\Tr[ a_i^\dagger a_j^\dagger] |0\rangle $, where the trace is over the $SU(N)$ indices $r,s$: $ \Tr[ a_i^\dagger a_j^\dagger] =  ( a_i^\dagger )_{r}^{~s} ( a_j^\dagger )_{s}^{~r}$. 
On the other hand, in the ungauged model we can have a state of the form $a_i^\dagger |0\rangle$. 
This state has energy $\mu/6$ at zero coupling. One can compute the first perturbative correction and we find that it is given by (see Appendix \ref{app:bmnper})
\be
E_{\rm adjoint} = {\mu \over 6 } +  { 9 \over 2 } { \lambda \over \mu^2 }  + \cdots 
\ee
where the dots are higher order terms in the $\lambda/\mu^3 $ expansion. The idea is that if we were to sum all the corrections and continue the answer to strong coupling we would get that $E_{\rm adj} \sim  
C \lambda^{1/3} $ for $\lambda/\mu^3 \gg 1 $. 

 It is interesting that one can obtain a relatively simple answer for this one loop shift for 
  the energy of  more general  adjoint states. This can be done using the general expression for the 
  one loop Hamiltonian in   
\cite{kim} and observing that it contains an additional piece for non-singlets in representation $R$ 
\beq
\Delta \hat{H}_\text{1-loop} = \Delta \hat{H}_\text{gauged, 1-loop} + \frac{9g^2}{2\mu^2} C_2(R)
\label{delta_e_main}
\eeq
The explicit expressions for $\Delta \hat{H}_\text{1-loop}$ and $\Delta \hat{H}_\text{gauged, 1-loop}$ are given in eq. (\ref{v_ns}). 

In general one should be careful with translating \nref{delta_e_main} to the energy shifts, since $\Delta \hat{H}_\text{gauged, 1-loop}$ may act differently on non-singlets.
This point deserves some clarifications. Since the trace is cyclic, operators $a^\da_i$, forming a single-trace singlet operator $\Tr [ a^\da_i a^\da_j \dots]$ are placed on ``a circle".
From this point of view, non-singlets have ``boundaries". So, generically, singlets and non-singlets have quite different Hilbert spaces and $\Delta \hat{H}_\text{gauged, 1-loop}$ may have
completely different eigenvalues. 

For example, in the gauged model on level two we have BPS states $\Tr[a^\da_i a^\da_j],\ i \neq j$. 
However, in the adjoint sector of the ungauged model operators $a^\da_i a^\da_j$ and $a^\da_j a^\da_i$ are different if $i \neq j$.
One can check explictly using eq. (\ref{v_ns}) that the state corresponding $a^\da_i a^\da_j + a^\da_j a^\da_i$ still does not receive corrections from $\Delta \hat{H}_\text{gauged, 1-loop}$, whereas
antisymmetric combination $a^\da_i a^\da_j - a^\da_j a^\da_i$ receives an additional shift of $4 g^2 \l \frac{3}{\mu} \r^2$. 
Nonetheless, both symmetric and anti-symmetric combinations get contribution from $\frac{9g^2}{2\mu^2} C_2(\text{adj}) = \frac{9g^2 N}{2\mu^2}$.
The reason the symmetric combination is still protected against $\Delta \hat{H}_\text{gauged, 1-loop}$ is that symmetrization restores the cyclic symmetry. It is natural to conjecture that
cyclic-symmetric non-singlets receive the same contribution from $\Delta \hat{H}_\text{gauged, 1-loop}$ as singlets.

%### [I THINK THIS IS LESS INTERESTING THAN THE ABOVE DISCUSSION. COMPUTATION BELOW DOES NOT TEACH US ANYTHING SO I HAVE COMMENTED IT OUT] 
%For example, one can study the non BPS  state $\sum_{i=1}^6 \Tr [ a^\dagger_i a^\dagger_i ]$ and its adjoint cousin $\sum_{i=1}^6 a^\dagger_i a^\dagger_i$.
%The singlet one has energy:
%\beq
%E^{K}_{singlet} = \frac{\mu}{3}+12 g^2 N \l \frac{3}{\mu} \r^2
%\eeq
%whereas the adjoint has an additional shift:
%\beq
%E^{K}_{adj} = \frac{\mu}{3}+12 g^2 N \l \frac{3}{\mu} \r^2 + \frac{9 g^2 N}{2\mu^2}
%\eeq
%where we used that $C_2({\rm adj}) = N$.

So far we have discused  the vacuum with $X=0$. One can also consider 
 a fuzzy sphere vacuum  (\ref{fuzzy_vacuum}) with  $X^a = \frac{\mu}{3} J^a,\ a=1,2,3$.
 In this case, one also expects $SU(N)$ Goldstone bosons. As in the discussion 
 in section \ref{LaXreg},  we could calculate the energy of states with non-trivial $SU(N)$ quantum numbers by considering states with $SU(N)$ angular momentum along the manifold spanned by the Goldstone bosons. We discuss this in more detail in Appendix \ref{SUNrot} where we found 
 a simple lower bound on the energy of the adjoint of the form 
\beq
\frac{g^2 N}{R^2}  =  { \lambda \over R^2 } \lesssim E_{\rm adjoint} ~,~~~~~~{\rm for} ~~~~ \mu^3 \gg \lambda 
\eeq
where $R^2 = \frac{1}{3N}{ \mu^2 \over 9 }  \Tr \l J_1^2+J_2^2+J_3^2 \r$ is the average radius of the fuzzy spheres.
This is consisent with the expectations based on \nref{PotAp}.

\subsection{Spectrum above the minimum} 

We now return to strong coupling. Around the thermal background we have argued that the minimum 
energy for the adjoint state is given in  \nref{Estimate}. We would now like to discuss excitations above these states. 
In the planar limit these excitations will be single strings of operators with a fundamental index at 
one end and an antifundamental at the other end, combined so that we 
an adjoint index in total. 
 We expect that these states would be strings whose ends are located in the 
high curvature region. 

Of course, when we quantize the string we expect a large number of modes. 
So we expect a number of energy eigenstates above the minimum given by \nref{Estimate}.
The first few are expected to be separated from the minimum by gaps which are of the same order of magnitude as the lowest energy itself \nref{Estimate}. In general, it seems complicated to determine this spectrum because it depends both on what is happening in the high curvature region as well as in the low curvature region. As the string gets more excited it can dip further into the region described by Einstein gravity. An example of an excitation would be a stretched folded string that goes from the high curvature region to some radial position $r_{\rm min}$. If $r_{\rm min}$ is within the weakly 
coupled region, then its motion could be as indicated in Figure \ref{string_tip}, namely the tip of the string goes into the weakly coupled region, it is slowed down by the string that pulls it from the boundary and it bounces back to the high curvature region\footnote{
We compute the phase shift for this motion in Appendix  \ref{TipMotion}. A similar computation for the single matrix model in the double scaling limit was done in \cite{Maldacena:2005hi,Fidkowski:2005ck} and matched to the matrix model.}. After it goes back into the high curvature region it could come back out with other worldsheet excitations, depending on the physics in the high curvature region. 
The full spectrum cannot be obtained unless we can solve both parts of the motion, namely the one in the low curvature region as well as the one in the high curvature region. In Appendix \ref{TipMotion} we discuss a toy problem where we assume that the string tip is reflected from the high curvature region without any further excitation, thought  this is probably not what happens in reality. 

When the excitation energy is large enough that the string can reach all the way to the horizon, something new happens. The string falls into the horizon and we end up with a string and an anti-string pair, each ending on the horizon. The minimum energy when this happens is given by the energy of a folded 
string that stretches all the way from the high curvature region to the horizon, 
\be \la{extf}
E_{\rm dec}  ={ 1 \over \pi }  \int_{r_0}^{r_{\rm high} \sim \lambda^{1/3} } d  r =  \tilde C \lambda^{1/3} - { r_0 \over \pi } = \lambda^{1/3} \left[ \tilde C - { 1 \over \pi } \left( { T 4 \pi \sqrt{d_0} \over 7 \lambda^{1/3} } \right)^{2 \over 5 } \right] 
\ee
where we expect that $\tilde C$ is an order one quantity bigger than $C$ in \nref{Estimate}. 

We will call this the ``deconfinement'' energy, because above this energy the adjoint is effectively behaving as two independent excitations, a quark and an anti-quark, corresponding to the string and antistring segments ending on the horizon.  
Furthermore, when a string ends on the horizon, there is an additional factor $N$ in the effective 
number of states. This arises as follows. When a string wraps the Euclidean black hole it has a disk topology, which produces and additional factor of $1/g_s \propto N$. This is {\it  in addition}  to the 
factor of $N$ that we get from all the possible values of the fundamental index at the boundary. This new factor is present for both the gauged
and ungauged models and it is related to the physics at the horizon.

When the string is not reaching the black hole horizon we can effectively think of the large $N$ Hilbert space as factorizing into the black hole part which lives in the singlet Hilbert space and non-singlet excitations that live close to the boundary.
 \beq
\mathcal{H}_{\rm non-singlets}  \sim  \mathcal{H}_{\rm singlets} \otimes \mathcal{H}_{\rm boundary ~ string  ~Fock~space }
\eeq
Furthermore there is a Fock space of boundary excitations, generated by the adjoint excitations which 
appear as strings with ends in the large curvature region, as in Figure \ref{fig:boundary}. Each generator has the degeneracy of an adjoint, or a factor of $N^2$.  ~\footnote{ We are idenfiying $N^2-1 \sim N^2$ since we are only discussing the leading $N$ effects.}

The strings that end on the horizon can be qualitatively viewed as extra tensor factors, one for the quarks and one for the anti-quarks (or strings or anti-strings), 
see Figure \ref{folded_string}(b). Each of these generates a Fock space. The string ending on the horizon is expected to have minimum energy 
$\tilde C \lambda^{1/3} /2$. 
The same is true for the anti-string. This is because the folded string whose tip is at the horizon has energy $\tilde C \lambda^{1/3}$ by definition. And this is becoming the string/anti-string pair. 
% \footnote{A string ending on the horizon is not a definite state, it has a relative large entropy associated to the fact that it is ending at the horizon. } 
The degeneracy of each generator also is proportional to $N^2$ but with a temperature dependent factor that can be computed by considering a string wrapping the black hole, which has an extra free energy given by \nref{extf} plus a logarithm of $N$, related to the factor of $1/g_s$ in the partition function mentioned above. This extra degeneracy is not exact, it simply reflects an  increase in the enropy of
the  combined black hole and string system, but we do not expect to be able to separate it cleanly into a black hole part and a string part. 
 We get the following {\it schematic} decomposition of the Hilbert space 
 \beq
\label{factorization}
\mathcal{H}_{\rm non-singlets}  \sim  \mathcal{H}_{\rm singlets} \otimes \mathcal{H}_{\rm boundary ~ string  ~Fock }  \otimes  \mathcal{H}_{\rm horizon~ string  ~Fock } \otimes  \mathcal{H}_{\rm horizon ~  antistring  ~Fock } 
\eeq

\subsection{The free energy} 

In this subsection we consider the free energy of the ungauged theory. Because gauging is removing of order $N^2$ degrees of freedom, and
given that the free energy is of order $N^2$, one might worry that the free energy of the gauged model would be very different than that of the 
ungauged one. 

In fact, large $N$ counting tells us that 
\be
 - \beta F_\text{ungauged } - \beta F_\text{gauged} = N^2 f(\lambda^{1/3} \beta ) 
\ee
 
For simplicity we could start considering the BMN model at weak coupling. In this case, in the ungauged theory we basically have $ 9 N^2$ bosonic harmonic oscillators, 
while in the gauged theory we have $ 8 N^2$ bosonic oscillators since the gauge constraint is essentially removing one matrix (the one we can diagonalize). 
The fermions give a subleading contribution in this high temperature limit. Therefore, in this case we get 
\be \la{WeakC}
f \sim - \log(\mu \beta ) ~,~~~~~~~~~~ \lambda \beta^3 \ll 1 ~,~~~~~~ {\lambda \over \mu^3 } \ll 1 
\ee
 
On the other hand, at strong coupling, $\lambda \beta^3 \gg 1 \gg  \beta \mu   $, 
where we can trust the black hole solution,  we have a different picture. 
The idea is that non-singlets are extra adjoint particles living  near the boundary of the gravity solution. Because of the factorization 
(\ref{factorization}) they contribute 
with   extra factors of the form 
\be
 (1 + N^2 d_{Adj} e^{ - \beta \lambda^{1/3} C } ) 
 \ee
 in the partition function. This is the contribution of just the lowest energy adjoint state and $d_{Adj}$ is its degeneracy. We expect it to be of order one. 
 The factor of $N^2$ comes from 
 the dimension of the adjoint representation. 
 Therefore we expect that the leading energy difference is
 \be
 \label{eq:prediction}
 \beta F_\text{gauged} - \beta F_\text{ungauged} = N^2 d_{Adj} e^{ - \beta \lambda^{1/3} C } ~,~~~~~~~~ \lambda \beta^3 \gg 1 
 \ee
 This shows that $F_\text{gauged}$ and $F_\text{ungauged}$ are exponentially close in 
 the strongly coupled limit, while they are different at weak coupling \nref{WeakC}.
 
 Let us emphasize that at strong coupling we have a reduction in the naively expected number of states in both theories. For that reason one might have 
 thought that the gauging or not gauging would have a large impact. 
 However, we see that this is not what is happening, both theories have a common low energy description. 
 
% As we have mentioned before, we essentially have a kind of factorization of the Hilbert space with a subsector which is the one giving rise to gravity and the low energy thermodynamics
 %and a decoupled set of excitations with an energy gap (\ref{Estimate}) where the non-singlets are residing. 
 %This sector consists of a tower of massive particles in the 
% adjoint and an approximate Fock space constructed from them. 
Using the factorized from of the Hilbert space, we can write a more precise form for the free energy difference
 \be
 f \sim \sum_{ n}  - (-1)^F \log \left[ 1 - (-1)^F e^{ - \beta E_{n} }  \right] 
 \ee
 where $n$ runs over all the adjoint states which can be bosons or fermions. This follows from standard large $N$ counting. 
 
 We can further improve the discussion 
  by including strings ending at the horizon. These contributions are most clear in Euclidean space. 
   They still give contributions to $f$ that are exponentially suppressed $\propto e^{ - \beta \lambda^{1/3}  \tilde C }$. These are smaller than 
 \nref{eq:prediction} because $C < \tilde C $.

 We have mentioned in the introduction that both ungauged and gauged models are
  unstable at very low temperatures. Here we will review more precise estimates for the decay rates
  (see eg. \cite{Horowitz:1997fr}). 
 Let us start from the gauged model. 
 Emitting a single D0 brane to infinity will lower the Bekenstein--Hawking entropy (\ref{BHentropy}). Therefore such process is  
 suppresed by:
 \beq
\label{eq:emission}
  P \sim \exp \l - { \partial  S \over \partial N }  \r = \exp \l - 2 { S \over N }   \r
  % \exp \l - 2N c_S T^{9/5}/\lm^{3/5} \r, \ c_S = 4^{13/5} 15^{2/5} (\pi/7)^{14/5}
 \eeq
 where $S$ is given in \nref{BHentropy}. The instability is unsuppressed when 
 \beq \la{TGL}
T_c \sim \frac{\lm^{1/3}}{N^{5/9}}
\eeq
Formally, at this temperature the dilaton becomes large at the horizon and one has to lift the gravity solution (\ref{geometry}) to 11d M-theory black string \cite{Itzhaki:1998dd}. 
Generically, black strings suffer from the Gregory--Laflamme  instability \cite{Gregory:1993vy}, which, in this case, also occurs at the temperature \nref{TGL}.  

The contribution from the lowest adjoint (\ref{eq:prediction}) will enchance (\ref{eq:emission}) by $\sim e^{ - \beta \lambda^{1/3} C }$. However at $T_c$ this factor is extremely
small $e^{-N^{5/9}}$. Therefore the instability in the ungauged model occurs at the same temperature. Indeed, as we have mentioned before, this instability is the instability of the black hole \textit{itself},
so excitations near the boundary should not affect it.

\section{Deconfinement  and the eigenvalues Polyakov loop holonomy }
\label{sec:hagedorn}

The main point of this paper is that in theories with gravity duals 
all non-singlets have high energies and are not dynamically important  at low energies. 
On the other hand, the arguments in \cite{Witten1998,Aharony2003} seem to suggest 
 that non-singlets are important for modifying the eigenvalue distribution of the Polyakov loop.  Furthermore, the fact that this distribution is not uniform 
is viewed as a signal of a black hole formation in the bulk. 

This seems to be in contradiction with what we are saying, since we are emphasizing that
the non-singlets are dynamically unimportant at low energies and strong coupling. 
 We will here show why there is no contradiction. 
  
  To start, let us 
suppose that we are studying the gauged model.
 Then the partition function includes the integral over the gauge field holonomy, which we can take to diagonal 
$U=\diag \l e^{i \th_1},\dots,e^{i \th_N} \r$. 
In the large $N$ limit it is convenient to introduce the normalized density function $\rho(\th)$
\beq
\int_{-\pi}^\pi d\th \ \rho(\th) = 1 
\eeq
and the corresponding moments 
$\rho_n = \int_{-\pi}^{\pi} d\th \ 
% \cos \l n \th \r 
e^{i n \theta} \rho(\th)$. 
The moments $\rho_n$ measure the non-homogenity of the density function.

Since we only have adjoint fields in the matrix model, 
 the energy can depend only on the relative distance between the
eigenvalues $\th_i-\th_j$. 
There is a constant repulsion of order one among eigenvalues $\th_i$ due to the group measure. Integrating out 
the matter fields leads to an attraction of eigenvalues.
At very low temperatures the repulsions dominates and the density function is uniform (in the 
BMN model). As the temperature increases, the attraction becomes stronger and stronger
 until the density
function jumps to a non-uniform distribution. In other words, eigenvalues from a cluster \cite{Aharony2003}.  
However, since the energy depends on the relative distance only, the absolute position of the cluster is not fixed, and one has to integrate over this zero mode. This is the reason why
the Polyakov loop in the fundamental representation is still zero after the transition.

This resembles the gravity computation of the Polyakov loop in the fundamental \cite{Witten1998}:
one can have a single string stretched between the horizon and infinity. Such a string has  a finite action and one could expect that the Polyakov loop will not be zero.
However, in the black hole background one has a normalizable mode of the  2-form $B_{\mu \nu}$, which couples to the string as 
\beq \la{Bint}
\exp \l i b \r ~,~~~~~~~b \equiv  \int B  
\eeq
And after the integration over $b$ one gets zero.

%This analogy suggests that the presence of a black hole is related to non-uniformness of the eigenvalue density. In our case, non-extermal D0 branes do form a black hole and so
%the eigenvalue density should be non-uniform. It is instructive to see how this statement is consistent with having a gap between singlets and non-singlets.
If we have just a single adjoint
particle of mass $C \lambda^{1/3}$ and degeneracy $d_{Adj}$, then the partition function reads as (see \cite{Aharony2003} for the derivation):
\beq \la{FirAtt}
Z = \int d \rho_1 \ \exp \l - N^2  |\rho_1|^2  \left[1-   d_{Adj} e^{-\beta C \lambda^{1/3}}  
\right] \r
\eeq
where in the exponent we have ignored the small terms proportional to $e^{-2\beta C \lambda^{1/3}},e^{-3\beta C \lambda^{1/3}}$ and so on.
The first term in brackets, the one, 
 comes from the $SU(N)$ measure, whereas the second term
  comes from the matter contribution, where $|\rho_1|^2$ is the contribution of the trace of the holonomy in the adjoint representation. 

Assuming that $d_{Adj}$ and $C$ are of order $1$ and $\beta \lambda^{1/3} \gg 1$, 
the integral is dominated by $\rho_1=0$.  Then, the density is uniform and we expect no black hole! 
This would be the right conclusion if the only states we had were the ones corresponding to strings with both ends at the high curvature region. For example, these are the only non-singlets around the gapped vacua of the BMN model in the strong coupling region. 

\begin{figure}[h!]
\centering
\includegraphics{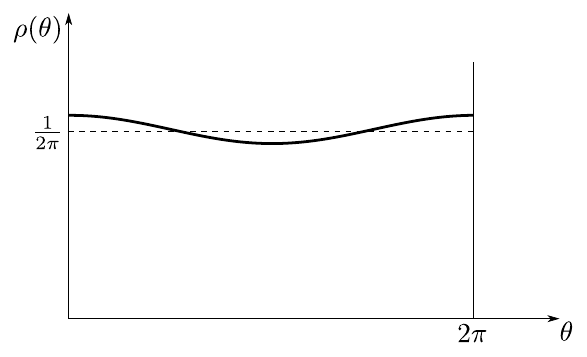}
\caption{The eigenvalue density of the Polyakov loop at strong coupling. It is only slightly non-uniform.}
\label{bump}
\end{figure}

However, apart from those strings, we can also have strings ending on the black hole. 
These strings effectively behave as quarks and antiquarks, with an overall constraint that there is 
an equal number of quarks and anti-quarks. We can view the integral over $b$ in \nref{Bint} as enforcing this constraint. 
Therefore we can now write a partition function of the form 
  %
% We propose the following resolution of this puzzle. Note that apart from the folded strings coming from infinity we also 
%have strings stretched between the horizon and infinity. However, because of the
%aforementioned Kalb--Ramond field such string will produce $\cos(B)\rho_1$, instead of $\rho_1$(which is simply the trace of the holonomy in the
%fundamental representation). Moreover, instead of naive $N$, we will have $N^2$ because the string ends on the horizon:
%So we now have additional contributions to the 
\beq
\label{z_string}
Z = \int d^2 \rho_1 d b \  \exp \l -N^2 \left[ 
 |\rho_1 |^2 -   d' e^{-\beta  \lambda^{1/3} \tilde C/2} ( e^{ i b} \rho_1 + e^{ - i b } \bar \rho_1 ) \right]  \r
\eeq 
where $d'$ is a temperature dependent quantity 
that is less important than the exponential factor we are 
explicitly writing. We will discuss the origin of $d'$ below.
We now see that, before integrating over $b$, 
the integral does have a non-trivial saddle point for $\rho_1$
\be
\rho_1^s = e^{ - i b} d' e^{-\beta  \lambda^{1/3} \tilde C/2 }
\ee
 Higher $\rho_n, \ n \ge 2 $ are suppressed by  factors of $( e^{-\beta \lambda^{1/3} \tilde C/2 })^n $.
It means that the density $\rho(\th)$ has a bump  determined by the cosine function, see 
Figure \ref{bump}. Of course, in this discussion we used the gravity solution to say what  answer to 
expect on the matrix model side. We have not derived this directly from the matrix model side!
 We are simply spelling out what answer we expect. 
 
We have a non-zero VEV of the Polyakov loop in the adjoint 
representation:
\beq \la{WadjE}
  \lll W_\text{adj} \rrr  =  \langle \Tr_{\rm adj} P e^{ i \int A } \rangle = N^2    |\rho^s_1|^2 \propto  N^2 e^{- \beta  \lambda^{1/3} \tilde C }
\eeq
In other words, the large energy required to stretch a string from the boundary to the horizon implies that the expectation value of this Wilson loop is very small. 
Of course,  the Polyakov loop in the fundamental is still zero since we have an integral over $b$.

Let us say a few words about the pre-exponent factor $d'$. The combination $d' e^{-\beta \lm^{1/3} \tilde C/2}$ in eq.~(\ref{z_string}) comes from a single string streched between
the boundary at $r=\tilde r_\infty$ and the horizon at $r=r_0$. Therefore,
\beq
d' \exp \l -\beta \lm^{1/3} \tilde C/2  \r =%
{\cal D} 
%\frac{1}{\sqrt{\det}}
 \exp \l - \beta \frac{\tilde r_\infty-r_0}{2\pi} \r
\eeq
The prefactor ${\cal D}$ arises from the one loop integral around the classical string configuration, which is a string that wraps the radial and Euclidean time directions. Due to the scaling properties of the 
solution (\ref{geometry}), it can only give a power law term in the temperature. 
The exponential term $\beta \frac{\tilde r_\infty-r_0}{2\pi}$ arises from   the classical string action. 
%Strictly speaking,
% we have a pre-exponent where $det$ is the determinant of string modes around the classical solution. 
%Generically, it is
%a complicated function of temperature and the coupling constant. However, one can use the hyperscaling symmetry of 
%the gravity solution (\ref{geometry}) to make the temperature to be an overall factor in front of the metric. 
%Therefore we expect that the determinant will be simply proportional to the temperature in some power. However, this 
%question deserves a separate study and below we are going to ignore the determinant.
As before, we expect that $\tilde r_\infty$ is a cutoff dependent quantity that is independent of the temperature, which we write as 
% Previously, for a folded string streched up to $r_\infty$ we put 
$\tilde r_\infty/\pi = \tilde C \lm^{1/3}$.
This constant is expected to be larger that $C$ in \nref{Estimate} since a string/anti-string pair ending on the horizon can decay into the 
massive string modes that live at the high curvature regions, which have the energy \nref{Estimate}. 
% However, since we do not know any details about the highly curved region,
%we do not have any reason to expect that for a single string $\tilde r_\infty = r_\infty$. We simply say that $\tilde r_\infty$ is of 
%order $\lm^{1/3}$: $\tilde r_\infty/2\pi = \lm^{1/3} \tilde C/2$. The entopy factor $d'$ is thus proportional to the length of the string in the bulk:
On the other hand the $r_0$ term gives a temperature dependent term in the exponent 
\beq
 \frac{\beta r_0}{2\pi} = \frac{\beta^{3/5}}{2\pi} \l \frac{4\pi \sqrt{\lm d_0}}{7} \r^{2/5}
\eeq
where we have used eq.~(\ref{eq:r0}) to find $r_0$ in terms of $\beta$.
%  This quantity was previously discussed in [ITZHAKI]

The adjoint particles that live near the high curvature region (see Figure \ref{fig:boundary}) contribute as $ e^{ - \beta \lambda^{1/3} C}$ to the expectation value in \nref{WadjE}, which is suppressed by $1/N^2$ compared to  \nref{WadjE}, but the exponential suppression is smaller, since $C< \tilde C$.  So the full expectation value in the adjoint is 
\beq \la{WadjFu}
  \lll W_\text{adj} \rrr   \propto  N^2 e^{- \beta  \lambda^{1/3} \tilde C }  + e^{ -\beta \lambda^{1/3} C }  + \cdots 
\eeq
where the dots refer to other contributions described by strings that are not ending at the horizon. The order $N^0$ contribution can be viewed
as arising from doing the Gaussian integral in \nref{FirAtt}.  In matrix model vacua with no black holes (such as the $X=0$ vacuum of the
BMN matrix model) we only get the second term in \nref{WadjFu}.

\section{ Further comments } 
 
 \subsection{Is there a bulk $SU(N)$  gauge field  associated to the $SU(N)$ global symmetry of the  ungauged model?  }
 \label{sec:confinement}
 
 The ungauged model has a global $SU(N)$ symmetry. According to the usual expectations, this should correspond to an 
 $SU(N)$ gauge symmetry in the bulk. 
 On the other hand, we have argued that the bulk theory, at least in the Einstein gravity region,  is essentially the same as that of
 the gauged model. 
  In our case,  the  states charged under $SU(N)$    are localized near the boundary of the 
 geometry.  The fact that a global symmetry might not extend over the full bulk is not at all surprising and it happens in other situations. 
 For example if we add $N_f$ massive fundamental fields, all with the same mass, 
  to an  $SU(N)$ gauge theory, then we have an $ SU(N_f) $ flavor symmetry. In the gravity dual, this is realized as  a brane 
 that reaches up to a finite distance $r_{\rm min}$ in the bulk \cite{Karch:2002sh}. 
 The larger the mass, the larger $r_{\rm min}$. For low energies, we explore the 
 bulk only in the region where $r< r_{\rm min}$ and we do not encounter states charged under the global flavor symmetry. 
 
Here something similar is happening, the bulk $SU(N)$ gauge symmetry, if present, is located only in the highly curved region, so it is not visible in the gravity region. And all bulk states that carry $SU(N)$ charge, have some excitations located in that highly curved region. 
    We can say that  the $SU(N)$ gauge symmetry we expected is ``confined'' in the bulk gravity region, but it is present in the highly curved 
     region.

 \subsection{Are there gauge fields on brane probes? } 
 
Let us consider the ungauged model. 
 Let us say that we have a probe D0 brane located in the region of the geometry described by Einstein gravity. 
 Does this brane probe have a gauge field on its worldvolume? Naively, one would say no, since we are dealing with the ungauged model. 
 On the other hand, we have argued that the bulk gravity region should be essentially the same for the gauged and ungauged models, so that we would expect a gauge field on the probe worldvolume. 
 
 We think that the right answer is the second, namely that there is a gauge field on the brane probes if the probes are in the Einstein gravity
 region, the region where $r< \lambda^{1/3}$.  This seems to be the only reasonable answer since these are the only kinds of D0 
 branes that we have in the ten dimensional string theory. 
 This gauge field imposes a constraint saying that 
   the number of strings ending on the D0 brane should be equal to the number coming out. 
   From the matrix model point of view,  
 the degrees of freedom on the brane probe are  effective low energy degrees of freedom that describe a complicated 
 bound state where the matrices have large fluctuations. Recall the discussion in section \ref{PolCal}.  For this reason they reflect more the dynamics of these degrees of freedom than the properties of the precise UV definition of the model. 
 
 On the other hand, if we consider a D0 brane probe in the highly curved region, which is described in perturbation theory, then we do not expect to have 
  a gauge field on the brane probe, since we do not have it in the ungauged model. 
 
 It would be interesting to understand what kind of transmutation the brane probe undergoes so that this happens as it crosses from the weakly curved bulk region to the strongly curved one.

 \subsection{The ungauged model and M-theory }
 
 In \cite{Banks:1996vh} the BFSS model was introduced as a tool to extract the    S-matrix for 11 dimensional M-theory. 
 In contrast to the discussion in most of this paper, the BFSS proposal is to consider a very low energy limit of this matrix model. 
 In this very low energy limit we go very deep inside the bulk, where the 11th dimension becomes large compared to other quantities and the physics is expected to reproduce the 11d one. 
 It seems that the difference between the gauged and the ungauged model is really lost when we go to such low energies, $E \propto 1/N$, 
  so that we could have as well started from the ungauged model also. 
 
 \subsection{Physical realizations } 
 
 It seems that the remarks in this paper suggest that if we wanted to build a quantum computer that simulates this problem we could start with a set of harmonic oscillators and Majorana fermions and then fine tune the interactions so that we get the ungauged model. 
 This seems simpler than producing the gauged model where the $SU(N)$ gauge redundancy  should emerge from some other further model. 
 In other words, it seems simpler to try to arrange for a model having an approximate $SU(N)$ global symmetry than having to produce one with 
 the $SU(N)$ gauge symmetry. 
 Because the energy of the non-singlets is higher than that of singlets we would expect that small perturbations that break the $SU(N)$ global symmetry should not be 
 important in the IR. Still, it is  important not to generate terms that lead to relevant perturbations of the model. But the number of those to fine tune seems smaller than those of all possible couplings. 
  
\section{Conclusions} 

We have seen that the Einstein gravity region can be present in both the gauged and ungauged versions of the model. 
The extra degrees of freedom of the ungauged model reside in the highly curved region of the geometry. 
We can say that both the gauged and ungauged models flow to the same theory in the infrared. Or that the ungauged model flows to the gauged model in the IR. 
Of course, it is not surprising that they have something in common, since the singlet sector is common to both theories. What we wanted to 
highlight here was that the non-singlets do not modify the gravity solution in the region where the gravity approximation is valid. 

A very similar story was found in the single matrix quantum mechanics in \cite{Gross:1990md}. There the two models coincided as long as the temperature was low enough. In that case, at temperatures higher than a critical temperature   the ungauged model would undergo a phase transition, somewhat reminiscent of the deconfinement transitions. See also \cite{Kazakov:2000pm} for a relation between that phase and black holes. 
In our case, the black hole phase is present both for the gauged and ungauged models. 
 
 We should emphasize that many of our statements can be rephrased in terms of expectations values of Wilson loops in the gauged model. We mainly talked about the 
 non-supersymmetric Wilson loop. For example,  a Wilson loop in the adjoint representation computed by a string like the one displayed in 
 Figure \ref{fig:boundary} (and extended along the time direction). This same loop has higher energy excitations where the string looks like the ones in 
 Figure \ref{folded_string}(b,c). 
 
 We have noted that the fact that the string has high tension implies that the eigenvalue distribution of the thermal holonomy, or Polyakov loop operator, 
 has only a very small inhomogeneity when we have black hole present, see Figure \ref{bump}.  One might have expected that the black hole formation would result in a stronger eigenvalue localization for the Polyakov loop.  This is the Polyakov loop of the full model, the UV theory, which is the only one we know how to define precisely in this theory. 
 
In  the Gurau-Witten tensor models, in a sense, the opposite from what we said here happens.
 In such models, in the leading large $N$ approximation the basic field behaves as a conformal field with low scaling dimension. Therefore we { \it do not have} 
 an energy gap to the non-singlets as we had in the D0 brane matrix model. In those cases the ungauged model seems a better starting point to describe the physics.

{\bf Acknowledgements } 

We have benefited from ongoing  discussions with E.~Berkowitz, 
M.~Hanada, E.~Rinaldi and P.~Vranas who  explained  to us the details and various results of their 
numerical simulations of the ungauged model \cite{Numeric}. Those 
 results were very helpful for us to gain confidence in the picture proposed here. 

We thank I. Klebanov and J. Polchinski  for discussions. 
J.M. is supported in part by U.S. Department of Energy grant
de-sc0009988 and the 
 % D.S. is supported by the
  Simons Foundation grant 385600.

\appendix
\section{Details of  the perturbative computations}
\label{app:bmn}
\subsection{Non singlets in the BMN matrix  model}
\la{app:bmnper}

In this Appendix we will study the BMN matrix model. 
Lagrangian reads as follows:
\beq
\begin{split}
\Lc=  \frac{1}{g^2} \Biggl( \oh \sum_{I=1}^9 \l \dot X^I \r^2 - \oh \l{\mu \over 3}\r^2 \sum_{a=1,2,3} \l X^a \r^2 - \oh \l{\mu \over 6}\r^2 \sum_{i \ge 4} (X^i)^2 + \oh \psi \dot \psi -
{\mu \over 8} \psi \gamma_{123} \psi + \\
-i {1 \over 3} \mu g  \sum_{a,b,c=1}^3 \Tr \l X^a X^b X^c \r \ep_{abc} + \frac{1}{4} \Tr \l [X^I,X^J]^2 \r + 
 i \oh \Tr \l \psi \ga^I [\psi, X^I] \r \Biggr)
\end{split}
\eeq
And supersymmetry transformations are given by:
\beq
\label{susym}
\begin{aligned}
\left[ Q \ep,X^I \right] &=  \psi \ga^I \ep(t) \\
\left[ Q \ep,\psi \right] &= \l \ga^I D X^I + c_I \mu X^I \ga^I \ga_{123}+ i \oh [X^I,X^J] \ga^{IJ} \r \ep(t) \\
[Q \ep,A]&= \ep(t) \psi \\
\ep(t) &= e^{-\frac{1}{12}\mu \ga_{123} t} \ep_0\\
c_a &=1/3 \ \text{for} \ a=1,2,3 \ \text{and} \ c_i=-1/6 \ \text{for} \ i \geq 4
\end{aligned}
\eeq
Note that supersymmetry transformations are time dependent.
In the supercharge we have an additional term proportional to $\mu$:
\beq
Q \ep =\Tr \l  -P^I \psi \ga^I \ep - i \frac{1}{2g^2} [X^K,X^L] \psi \ga_{KL} \ep -  \frac{\mu}{g} c_I X^I \psi \ga^I \ga_{123} \ep  \r
\eeq
Apart from the gauge transformation generator, supercharge anticommutator now also contains rotations generators $M_{\al \bt}$:
\beq
\label{qqm}
\{Q_\al,Q_\bt\} = 2H \delta_{\al \bt} + 2 \Tr \l G X^L \r \ga^L_{\al \bt} + M_{\al \bt}
\eeq
\beq
\begin{split}
M_{\al \bt} = - (\mu/3) \sum_{i,j \ge 4} 
\Tr(X^j P^i) \l \ga_{ji} \ga_{123} \r_{\al \bt} +
 (2/3) \mu \sum_{a,b,c=1}^3 \Tr(X^a P^b) \ep^{abc} \ga^c_{\al \bt} \\
-\cfrac{\mu}{6g^2} \sum_{i,j \ge 4} \Tr \l  \psi \ga_{ij} \psi \r  \l \ga^{i j} \ga^{123} \r_{\al \bt} 
+\cfrac{\mu}{12g^2} \sum_{a,b \in 1,2,3} \Tr \l  \psi \ga_{ab} \psi \r  \l \ga^{ab} \ga^{123} \r_{\al \bt} 
\end{split}
\eeq
Also recall that the gauge transformation generator is given by:
\beq
G = \frac{i}{2g^2} \l  2 [D_tX^I,X^I] + [\psi_\al,\psi_\al] \r
\eeq
Again, since the super charge gauge is invariant:
\beq
[Q_\al,G]=0
\eeq
Hamiltonian is given by:
\beq
\label{hamm}
\begin{split}
H= \frac{1}{g^2} \Tr \Bigg( g^4 \cfrac{P_I^2}{2}- \frac{1}{4} [X^I,X^J]^2 - i \oh \psi \ga^I [\psi,X^I] + \oh \l \frac{\mu}{3} \r^2 \sum_{a=1,2,3} \l X^a \r^2 + \\
+  \oh \l \frac{\mu}{6} \r^2 \sum_{i \ge 4} (X^i)^2 +
{\mu \over 8} \psi \gamma_{123} \psi +
 i {\mu \over 3}  \sum_{a,b,c=1}^3 \Tr \l X^a X^b X^c \r \ep_{abc} \Bigg)
\end{split}
\eeq

However since the supersymmetry transformations are time-dependent now, commutator of Hamiltonian with a supercharge is
proportional to a supercharge:
\beq
[Q_\al,H]=- \Tr (\psi _\al G)-\cfrac{\mu}{12} Q_\bt \ga^{123}_{\bt \al}
\eeq

As in the BFSS case, we can remove the gauge transformation generators from the SUSY algebra by imposing (\ref{save}) and redefining 
Hamiltonian:
\beq
\label{bmn_new}
H^\text{new}=H- \Tr \l X^1 G \r
\eeq
%where $H^t_\al$ is given by:
%\beq
%\begin{split}
%H^t_\al =  -\cfrac{1}{12} \mu \sum_{i=1}^9 \Tr \l (\psi \ga^i \ga_{123})_\al P^i \r + \cfrac{1}{18} \mu^2 \sum_{i=1,2,3} 
%\Tr \l X^i (\psi \ga^i)_\al \r -  \cfrac{1}{36} \mu^2 \sum_{i=4}^9 \Tr \l X^i (\psi \ga^i)_\al \r - \\
% \cfrac{5}{3} i g \mu
%\sum_{i,j,k=1,2,3} \ep_{ijk} \Tr \l (\psi \ga^i)_\al X^j X^k \r - 
%\cfrac{1}{6} i g \mu \sum_{k,l=4 \ \text{or} \ k=1,2,3}^9 \Tr \l (\psi \ga_{123} \ga_{kl})_\al [X^k,X^l] \r
%\end{split}
%\eeq
Now lets discuss the perturbative spectrum of this model.
It would be convenient to introduce indices from the beginning of the Latin alphabet $a,b,c,\ldots$  running 
from 1 to 3 whereas $i,j,k,\ldots$ run from 4 to 9.
We can introduce creation-annihilation operators by:
\beq
\begin{split}
a_b = \sqrt{\cfrac{3}{\mu}} \l \cfrac{g P_b}{\sqrt{2}} - i \cfrac{\mu}{3 \sqrt{2} g} X_b \r \\
a_i = \sqrt{\cfrac{6}{\mu}} \l \cfrac{g P_i}{\sqrt{2}} - i \cfrac{\mu}{6 \sqrt{2} g} X_i \r \\
\end{split}
\eeq
$SO(6)$ sector oscillators has mass $\mu/6$ and $SO(3)$ sector has mass $\mu/3$. Free Hamiltonian reads as:
\beq
H_0 = \cfrac{\mu}{3} \Tr a^\da_b a_b + \cfrac{\mu}{6} \Tr a^\da_i a_i
\eeq
Let us concentrate on the lightest $SO(6)$ sector. 
The leading order correction to the energy was computed in \cite{kim} to be:
\beq
\label{v1eff}
V^{(1)}_\text{eff}=g^2 \l \cfrac{3}{\mu} \r^2 \l N :\Tr a^\da_i a_i: + \cfrac{1}{2} :\Tr [a^\da_i , a_i][a^\da_j,a_j]:
-\cfrac{1}{2}:\Tr [a^\da_i,a_j][a^\da_i,a_j]:-:\Tr [a^\da_i,a^\da_j][a_i,a_j]:  \r 
\eeq
where one has to sum over all possible indices $i,j$ ranging from 4 to 9.

Therefore for the simplest adjoint state $a^\da_i|0\rrr$ the first-order correction is positive. 
We can rewrite the effective potential in a bit different form \cite{kim}:
\beq
\label{v_ns}
V^{(1)}_\text{eff}=g^2 \l \cfrac{3}{\mu} \r^2 \l \cfrac{1}{2} \l : \Tr [a^\dagger_i,a^i] T^a : \r  \l : 
\Tr [a^\dagger_i,a^i] T^a :  \r
-\cfrac{1}{2}:\Tr [a^\da_i,a_j][a^\da_i,a_j]:-:\Tr [a^\da_i,a^\da_j][a_i,a_j]:  \r 
\eeq
The last two terms are exactly the 1-loop dilatation operator in $\mathcal{N}=4$ SYM. One can show that the first term is zero
for singlet states. For non-singlet states build from $a^\dagger_i$ its value is proportional to the number of non-contracted
indices. That is, this term is proportional to the quadratic Casimir of the corresponding representation. To sum up, at
1-loop level the energy of non-singlets in the representation $R$ goes up:
\beq
\Delta \hat{H}_\text{1-loop} = \Delta \hat{H}_\text{gauged, 1-loop} + \frac{9g^2}{2\mu^2} C_2(R)
\label{delta_e_app}
\eeq
However, if we study
the modified Hamiltonian (\ref{bmn_new}), we have to take into account the correction coming from the 
operator $\Tr \l X^1 G \r$. Second-order perturbation theory for this additional correction
yields:
\beq
-g^2 \l \cfrac{3}{\mu} \r^2 \l  : N \Tr a^\da_i a_i : + \cfrac{1}{2} :\Tr [a^\da_i , a_i][a^\da_j,a_j]:  \r
\eeq
This contribution completely cancels the non-singlet contribution in (\ref{v_ns}). It means that the theory with supersymmetric Wilson loop has 
a protected $SO(6)$ sector, like the original theory. For example, the energy of the simplest adjoint state $a_i^\dagger | 0 \rrr$ is protected and is given by
\beq
E=\cfrac{\mu}{6}
\eeq

\subsection{BFSS model}
\label{sec:bfss_pert}
In this Appendix we will discuss the perturbaive spectrum of adjoints and derive the estimate (\ref{PotAp}). 
We consider a backround with diagonal matrices $\lll X^I \rrr = B^I = \diag(B^I_1,\dots,B^I_N)$. These break the $SU(N)$ symmetry, so we will have a compact manifold of 
 Goldstone bosons. In principle we need to study the quantum mechanics on this manifold. Since
  $SU(N)$ acts on this manifold this quantum mechanics gives rise to states charged under $SU(N)$. 
  From the analysis of the gauged model we know that there is a single uncharged state. We now want to discussed the states with   $SU(N)$ charges.  
   One can obtain their spectum
as follows. 

We want to study the angular motion around the diaginal background $\lll X^I \rrr = B^I$. 
Therefore we focus on $X^I$ in the following form:
\beq
\label{u_rot}
X^I(t) = U(t) B^I U^\dagger(t)
\eeq
and plug this expression into the original Lagrangian to find the effective action for $U$:
\be
\label{u_eff}
S = { 1 \over 2g^2 } \int dt \Tr \l (U^\dagger \partial_t U )_r^{~s} ( \vec B_r - \vec  B_s )^2  (U^\dagger \partial_t U )_s^{~r} \r
\ee
where we have used a shot-hand notation $\sum_{I=1}^9 (B^I_r-B^I_s)^2= \dist{r}{s}^2$

Now we need to analyse the symmetries carefully. Under the original $SU(N)$ 
gauge transformation $L$, $X^I$ transforms as in eq. (\ref{u_rot}): $X^I \ra L X^I L^\dagger$. It is equivalent to
multiplying $U$ by $L$ from the \textit{left}:
\beq
U \ra L U 
\eeq
In other words, the original gauge group $SU(N)$ acts by left rotations of $U$. Obviously, it is a symmetry of 
(\ref{u_eff}). So the states will come in $SU(N)$ multiplets.
%So we can keep $B^I$ fixed and rotate $U$ instead. If we look at $B^I$ as a rigid body in the $SU(N)$ space, then 
%it means that the original $SU(N)$ gauge charges correspond to angular momenta in the frame where the body is fixed.
The corresponding charges are given by:
\beq
G_r^{~s} = \frac{1}{g^2} \Tr \l U^\dagger \partial_t U  [B^I,[B^I,U^\dagger T_r^{~s} U]] \r 
\eeq
It is straightforward to check that they coinside with the charges $G$ in the Gauss law (\ref{Gauss}), as expected. 
Note that $G_r^{~s}$ is not a matrix element, but a charge corresponding to $SU(N)$ algebra generator $T_{r}^{~s}$ which has only one non-zero element 
on $r$-row and $s$-column.
% The expression for the charges implies that the diagonal elements  $G_s^{~s}  =0$ vanish (no sum). 

However, we can also multiply $U$ by a $SU(N)$ matrix $R$ from the \textit{right}:
\beq
U \ra U R
\eeq
This is not a symmetry of (\ref{u_eff}). So the corresponding current
\beq \la{tildeCha}
\widetilde G_r^{~s} = \frac{1}{g^2} \Tr \l U^\dagger \partial_t U  [B^I,[B^I,T_r^{~s} ]] \r = \frac{1}{g^2} 
  (U^\dagger \partial_t U)_r^{~s} \dist{r}{s}^2 
\eeq
does not commute with the Hamiltonian. Nonetheless, as was clarified in \cite{Marchesini:1979yq,Gross:1990md} left and right multiplications
of $U$ are tightly related. To understant this, let us consider a wave function $\Psi^a_\Rc (X^I),\ a=1,\dots,\dim \Rc$ 
in some representation
$\Rc$ under the gauge group. Since it lives in the representation $\Rc$ it has, by definition, the following decomposition:
\beq
\Psi^a_\Rc(X^I) = \sum_{b=1}^{\dim \ \Rc} U_\Rc^{ab} \psi_b\l B^I \r
\eeq
where $U^{ab}_\Rc$ is the $ab$ matrix element of $U$ in the representation $\Rc$. We are interested solely in $\psi_b\l B^I \r$
which also lives in $\Rc$.
Left $SU(N)$ rotations of $U$ rotate $\Psi^a_\Rc(X^I)$ and $U$, leaving $\psi_b(B^I)$ invariant. 
Whereas right rotations
transform $U$ and $\psi_b(B^I)$, leaving $\Psi^a_\Rc(X^I)$ invariant. Note that in both cases the representation $\Rc$
\textit{is the same}. It means that charges $\wt G_r^{~s}$ act on states $\psi_b\l B^I \r$ by the corresponding 
generator $(T_r^{~s})_\Rc$ in the representation $\Rc$.

What is the physical meaning of operators $\wt G_r^{~s}$? One can think about $X^I$ as a rigid body in a space acted on by the
$SU(N)$ transformations. Since  $\Psi^a_\Rc(X^I)$ stays invariant under $\wt G_r^{~s}$, they have a meaning of
angular momentum operators in the frame where the body is fixed. It is well-known from the classical mechanics,
that such operators are very useful for studying the rigid body motion, despite the fact that they are not conserved.

As we have just mentioned, $\tilde{G}_r^{~s}$ do not commute with the Hamiltonian. However, the 
Hamiltonian can be expressed in terms of them. Indeed, it is easy to see that
\be  \la{PotNs}
H =  \frac{g^2}{2} \sum^N_{r,s=1} { \wt G_s^{~r} \wt G_{r}^{~s} \over ( \vec B_s - \vec  B_r )^2 } 
\ee

When we focus on a particular repsentation $\Rc$, then $\wt G_r^{~s}$ act by the $SU(N)$ generator $(T_r^{~s})_\Rc$ in this
repsentation. The corresponding wave function depends only on $B^I$. This wave function is exactly what we
previously called $\psi_b(B^I)$. Naively, $H$ is a $\dim \Rc \times \dim \Rc$ matrix. However, 
the expression for the charges in \nref{tildeCha} implies that the diagonal elements  ${\tilde G}_s^{~s}  =0$ vanish (no sum). 
 This implies that we have 
 much less components. As we will see shortly, in the simplest case of the adjoint
representation instead of the naive $N^2-1$ we will have only $N-1$ states. Generically all these states
have different energies. However, we would like to emphasize that
each of these $N-1$ eigenstates has a degeneracy $N^2-1$(or $ \dim \Rc $ in the generic case) 
because of the angular degree of freedom $U$ which we have eliminated. 

So far we have been using the canonical quantization of non-singlets. Below we will re-derive (\ref{PotNs})
using the path integral techniques. Moreover, the fact that we always have a degeneracy $\dim \Rc$ will
become especially clear.

For the case of a single matrix model the result (\ref{PotNs}) was obtained in \cite{Marchesini:1979yq,Gross:1990md,Boulatov:1991xz}. 
But unlike the one matrix case, we cannot diagonalize  all the matrices simultaneously for  generic matrix configurations. 
Therefore (\ref{PotNs}) will receive higher loop corrections from off-diagonal fluctuations of $X^I$ and $\psi_\alpha$.

There is another, more clean-cut way, how to derive eq. (\ref{PotNs}) which will illustrate the above points. 
As we have mentioned in section \ref{sec:Wilson} if we are interested in excitations of the \textit{ungauged} model in a representation $\bar{\Rc}$(conjugate to $\Rc$) 
under $SU(N)$ we can study
the \textit{gauged} model coupled to a Wilson line in representation $\Rc$:
\beq
\label{aux_W}
\dim \Rc \Tr_\Rc P \exp \l i \int dt \ A_t \r = \dim \Rc \Tr P \exp \l i \int dt \ (A_t)_r^{~s} (T_{s}^{~r})_\Rc \r
\eeq
At this point it is by no means necessary to think about $A_t$ as a gauge field. In the ungauged case one can think about it as an auxillary Lagrange multiplier which forces the states to live
in a particular representaion. Note, however, that in the ungauged model we have to multiply the Wilson loop by the dimension
of the corresponding representation. This can be explained as follows. In the ungauged model we put $A_t$ to be zero.
We can achieve this by inserting the delta function into the path integral:
\beq
\delta \l P \exp \l i \int dt \ A_t \r \r 
\eeq
Now we can re-express the delta function in terms of characters \cite{Boulatov:1991xz}:
\beq
\delta \l P \exp \l i \int dt \ A_t \r \r = \sum_\Rc \dim \Rc  \Tr_\Rc P \exp \l i \int dt \ A_t \r
\eeq

We separate $X^I$ into the constant background $B^I$ and a 
 fluctuation $Y^I$: $X^I = B^I + Y^I$. Then the part of the (bosonic) action containing $A_t$ reads as:
\beq
\frac{1}{2g^2} \int  dt \ \Tr \l \pr_t Y^I + i[A_t,Y^I] + i[A_t,B^I] \r^2
\eeq
At 1-loop level we can simply ignore $Y^I$ and integrate out only $A_t$. However at higher loops one has to take $Y^I$ into account. Without $Y^I$ we have a simple quadratic action for $A_t$:
\beq
\label{quad}
-\frac{1}{2g^2} \int dt \ \dist{r}{s}^2 (A_t)_r^{~s} (A_t)_s^{~r}
\eeq 
Overall, we have the following expression:
\beq
\dim \Rc \int D A_t \ \exp \l -\frac{1}{2g^2} \int dt \ \dist{r}{s}^2 (A_t)_r^{~s} (A_t)_s^{~r}  \r \Tr P \exp \l i \int dt \ (A_t)_r^{~s} (T_{s}^{~r})_\Rc \r
\eeq
Integration over $A_t$ yields the angular potential (\ref{PotNs}). 

Note that because of the factor $\dim \Rc$, each eigenstate of the angular potential (\ref{PotNs}) will contribute to the
partition function with degeneracy  $\dim \Rc$.

In the adjoint case $(\wt G_r^{~s})_\Rc$ acts by a commutator with $T_{r}^{~s}$ on the matrix $w$ in the $SU(N)$ algebra. Moreover $w$ has to be diagonal, since the diagonal charges $
\widetilde{G}_s^{~s}$ vanish:
$w = \diag(w_1,\ldots,w_N)$. 
So we have the following eigenvalue problem\footnote{This can be easily obtained using the following relations $[ T_{r}^{~s},w] =(w_s-w_r) T_{r}^{~s}$, 
$T_r^{~s} T_r^{~s} = 0,\ r \neq s$ and $(T_r^{~s} T_s^{~r})_{vo} = \delta_{rv} \delta_{ro}$ (no sum over $r,s$)}
\beq
\label{e_adj}
E w_r =\frac{g^2}{2} \sum_{s=1, ~s \neq r}^N \frac{w_r - w_s}{\dist{r}{s}^2} ~,~~~~~~~{\rm with} ~~~ \sum_{r=1}^N w_r=0
\eeq
where the last constraint comes from the restriction that the diagonal matrix $w$ is in the adjoint. 
 This Eigenvalue problem will have $N-1$ eigenstates. We could identify the potential in figure \ref{potential} as the lowest energy state of this 
 Hamiltonian, as a function of the $\vec B_s$. 
 In general, the eigenvalues will depend on the particular pattern of the distances $(\vec B_r - \vec B_s)^2$. 
 In the next subsection we solve it for the case when a large number of vectors $\vec B_s$  is uniformly distributed on $S^8$.

There is another very simple case when the energy can be obtained exactly.
Suppose we are considering a configuration where the  $N$ vector $\vec B_s$ take only two values:
 $N_1$  are given by  $\vec{B}_1$ and the rest, $N_2=N-N_1$,  by $\vec{B}_2$.  
Equation (\ref{e_adj}) will be well-defined if $w_i$ obey the same property: there are $N_1$ coordinates $w_1$ and $N_2$ of $w_2$.
Using the constraint $N_1 w_1+N_2 w_2 = 0 $ we easily obtain the energy:
\be
E=\frac{\lm}{2\dist{1}{2}^2}
\ee
The result depends only on the sum $N_1 + N_2$ and not on the individual $N_1$,$N_2$. Also, the factor $N$ in the numerator is important: we expect that the energy of 
adjoint excitations will scale as $\lm^{1/3}$. Indeed, the expected size of the ground state wave function is $X \approx \lm^{1/3}$. 
This is the value of $X$ where this computation breaks down. We can then 
 identify the energy at this  value of $X$ as the order of magnitude of the energy of the adjoint excitation
%  then the mass of adjoint excitations:
\be
\label{adj_est}
E_\text{adj} \approx \lm/X^2 = \lm^{1/3}
\ee

\subsubsection{ Solving the potential for a uniform distribution }

There is another case when we can solve (\ref{e_adj}) exactly. Namely, lets consider the large $N$ limit with the $N$ vectors $\vec B_s$ 
uniformly distributed on $S^8$ of radius $X$. 
% In the Appendix \ref{sec:bfss_pert} 
We will show below  that the lowest energy state has energy
\beq
E_1 = \frac{9 \lm}{28 X^2}
\eeq
%and that the  highly excited states have the following spectrum:
%\be
%\label{n_spectrum}
%E_n = \frac{3 \lm}{8X^2} \l 1-\frac{1}{7} \l \frac{2}{8! n} \r^{3/4} \r,\ n \gg 1
%\ee
and we will further compute spectrum around the ground state. 
%The upshot of these calculations is that there is a gap of order $\lm/X^2$ between singlets and adjoints. Moreover, the level spacing between adjoints is of the same order.
%In the following section we will compare this spectrum for large $n$ with the gravity prediction. Suprisingly, it has the same form as $(\ref{n_spectrum})$, except that instead of 
%$n^{3/4}$ one has $n^{2/3}$.

% ------------

This is shown as follows. 
With a  large number of vectors    uniformly distributed on $S^8$ we can make a continous approximation. Then 
eq. (\ref{e_adj})
becomes
\be
\label{cont_adj}
E w(\vec n ) = \frac{\lm}{2X^2 \text{Vol}_{\text{S}^8}} \int d\Omega'_8  \ \frac{w(\vec n)-w(\vec n')}{|\vec{n}-\vec{n}'|^2},~~~~~~ \ \int d\Omega_8  \ w(\vec n) = 0
\ee
where $\vec n $ and $\vec n'$ belong to a $S^8$ of unit radius.
Now the adjoint problem (\ref{cont_adj}) has $SO(9)$ rotation symmetry. It means that the eigenfunctions are basically given by the spherical harmonics in nine dimensions and the energy
depends only on the total angular momentum $l$. 
It is the most convenient to evaluate the energy using the wave function which depends only on one polar angle $\th$ 
(the angle between the unit vector and $X^9$ axis). 
For such functions the measure $dy$ reads as $\text{Vol}_{\text{S}^7} \ \sin^7{\th} d\th = \text{Vol}_{\text{S}^7} \ (1-t^2)^3 dt$.
In this case $w(\vec n)$ is simply the Gegenbauer polynomial $C^{(7/2)}_l(t)$. Therefore,
\be
E_l C^{(7/2)}_l(1) = \frac{35 \lm}{64 X^2} \int_{-1}^{1} dt \ \frac{(1-t^2)^3}{2(1-t)} \l C^{(7/2)}_l(1)- C^{(7/2)}_l(t) \r
\ee
which leads to the following energies:
\beq
\label{l_spectrum}
E_l = \frac{3 \lm }{8X^2}\l 1- \frac{1}{C^{(7/2)}_l(1)} \r \sim \frac{3 \lm }{4X^2}  \l 1- \frac{1}{7 l^6} \r, \ l \gg 1
\eeq
This energy comes with a degeneracy 
\be
N_l = \cfrac{(2l+7)(l+6)!}{7! l!} \sim \cfrac{2l^7}{7!}, l \gg 7
\ee 
%Rewriting (\ref{l_spectrum}) in term of the uniform spectral density
%$dn = N_l dl$ we obtain (\ref{n_spectrum})  
%  the following spectrum
 %for large $n$.
%\be
%E_n = \frac{3 \lm}{8X^2} \l 1-\frac{1}{7} \l \frac{2}{8! n} \r^{3/4} \r,\ n \gg 1
%\ee
% Which is eq. (\ref{n_spectrum}) 

%	I THINK IT IS MISLEADING TO WRITE IT AS IN (\ref{n_spectrum})  BECAUSE IT SUGGESTS THAT $n$ IS QUANTIZED. I UNDERSTAND THAT YOU THINK OF IT A DETERMINING A DENSITY OF STATES RATHER THAT A SPECTRUM. I THINK THAT 
%	 (\ref{l_spectrum}) WOULD BE ENOUGH AS A FINAL ANSWER. 

It is interesting that we get a finite range of energies, from a minimum one to a maximum. This pattern is similar to what we get by a 
simple WKB quantization of a toy model for the motion of a folded string in Appendix \ref{TipMotion}. However, in the gravity case, we can also have
the possibility of the string falling into the black hole which leads to a much larger number of states, a number proportional to $N^2$, one 
factor of $N$ each for the separate  string and anti-string ending on the black hole.

\subsection{Goldstone modes and $SU(N)$ rotators for the BMN model vacua } 
\la{SUNrot}

Now let us discuss the spectrum   around other  vacua,  where the matrices have non-zero expectation values of the form 
 $X^a=\frac{\mu}{3} J^a$. This case can be analyized as in the previous section.
The only difference is that 
the initial action is 

\beq
\label{s_prelim}
S = \frac{\mu^2}{18 g^2} \int dt \ \sum_{a=1}^3  \Tr  [J^a,U^\dagger \pr_t U ]^2 =  \frac{\mu^2}{18 g^2}  \int dt \ \sum_{a=1}^3  \Tr U^\dagger \pr_t U [J^a,[J^a,U^\dagger \pr_t U ]] 
\eeq

And the right $SU(N)$ charge equals:
\beq
\wt G^p=\frac{\mu^2}{9 g^2}  \int dt \ \sum_{a=1}^3  \Tr \l  [J^a,[J^a,U^\dagger \pr_t U ]]  T^p \r
\eeq
with $T^p, \ p=1,\dots,N^2-1$ belonging to $SU(N)$ algebra.

As we have mentioned in the main text, $J^a$ are not neccessary in the irreducible representation. 
Generically, we need to decompose it into $L$ irreducible representations of dimensions 
$N_k, \ k=1,\dots,L$ such that
$N_1 + \ldots + N_L = N$. For simplicity we study the maximal representation $L=1$ and $N_1=N$, although the calculation below can be generalized to $L>1$ case.
Even for the maximal case, when we have only one representation it is quite difficult to obtain the exact spectum. However, it is easy to find a sensible \textit{lower bound}.

Since we have only the kinetic term the energy equals:
\beq
\label{gold_e}
E  = \frac{\mu^2}{18 g^2}  \Tr \int dt \ \sum_{a=1}^3 [J^a,U^\dagger \pr_t U ]^2
\eeq

Generically, there are many ways to select Lie algebra generators $T^p$. However, there is a very special choice of $T^p$, namely the fuzzy spherical harmonics $Y^j_m,\ j=1,\dots,N-1,\ m=-j,\dots,j$. 
The nice thing about them is that they are eigenvalues of the fuzzy sphere Laplacian:
\beq
\label{fuzzy_lap}
\sum_{a=1}^3 [J^a,[J^a,Y^j_m]] = j(j+1) Y^j_m
\eeq
Also they are orhogonal:
\beq
\Tr \l  Y^{j}_m Y^{j'}_{m'}\r = \oh \delta_{j j'} \delta_{-m m'}
\eeq
Because of that they also satisfy the completness relation:
\beq
\label{useful2}
\sum_{jm} \l Y^j_m \r_{r}^{~s} \l  Y^j_{-m} \r_{v}^{~o} = \oh \l \delta_{r}^o \delta_{v}^s - \frac{1}{N} \delta_{r}^o \delta_{v}^s  \r,\ r,s,v,o=1,\dots,N
\eeq

Correspondingly we have the non-conserved  charges $\wt G^{j}_m$:
\beq
\wt G^j_m = j(j+1) \cfrac{\mu^2}{9g^2} \Tr \l U^\dagger \pr_t U Y^j_m \r
\eeq
Finally, we can rewrite the Hamiltonian in terms of $\wt G^j_m$ using eq. (\ref{useful2}):
\beq
\begin{split}
H =  \frac{\mu^2}{9 g^2} \sum_{a=1}^3 \sum_{jm} \Tr \l U^\dagger \pr_t U Y^j_m \r \Tr \l Y^j_{-m} [J^a,[J^a,U^\dagger \pr_t U]] \r = \\
= \frac{\mu^2}{9 g^2}  \sum_{jm} j(j+1) \Tr \l U^\dagger \pr_t U Y^j_m \r \Tr \l Y^j_{-m} U^\dagger \pr_t U \r = \\
= \frac{9g^2}{\mu^2} \sum_{jm} \frac{\wt G^j_m \wt G^j_{-m}}{j(j+1)}
\end{split}
\eeq
As we have promised, we have re-expressed the Hamiltonian in terms of charges $\wt G$. 
If we focus on some particular representation $\Rc$, then $\wt G^{j}_m$ act as
Lie algebra generators $(Y^j_m)_\Rc$ in this representaion. For example, the sum 
$\sum_{jm} \wt G^j_m \wt G^j_{-m} = C_2(\Rc)$ equals to the 
quadratic Casimir of the representation. Since $j \le N-1$ we obtain the following lower bound for the energy:
\beq
H \ge \frac{9g^2}{\mu^2} \frac{1}{N(N-1)} C_2(\Rc).
\eeq

The above derivation can be repeated when we have several fuzzy spheres with corresponding representations $N_k$. In this case one arrives at the following bound:
\beq
\label{fuzzy_bound}
H \ge \frac{1}{\text{max} \ N_k(N_k-1)} \frac{9g^2}{\mu^2} C_2 \l \mathcal{R} \r
\eeq
%Recall that in the original model all the states are in the adjoint representation of $SU(N)$. Therefore we can not choose $\mathcal{R}$ to be arbitrary, it should stem
%from the adjoint of $SU(N)$. 
For the adjoint representation of $SU(N)$ the quadratic Casimir $C_2(adj)$ is simply $N$.
The other representations that appear are those that can arise from products of adjoints. These are the representations that transform trivially under the $Z_N$ center
of $SU(N)$.

For ``small'' fuzzy spheres, when $N_k \sim \mathcal{O}(N^0)$ and $L \sim \mathcal{O}(N)$, $E_\text{adj} \gtrsim \lambda/\mu^2$. However, if we have
a ``big'' sphere, when some $N_k \sim N$ and so $L \sim 1$, adjoints can have much smaller energy $E_\text{adj} \ge \frac{g^2}{N \mu^2}$.
Note that both these bounds are consistent with
\beq
\label{fuzzy_estimate}
E \sim \frac{g^2 N}{R^2} C_2(\mathcal{R})
\eeq
with $R^2 = \frac{1}{3N} \Tr \l J_1^2 + J_2^2+J_3^2 \r$.
Since for each irreducible representation $N_k$ we have the following identity:
\beq
J_1^2 + J_2^2 + J_3^2 = \frac{N_k^2-1}{4} \ {\bf 1}
\eeq
where the right hand side is simply the quadratic Casimir of $SU(2)$ in the representation of dimension $N_k$.

\nref{fuzzy_estimate} is what we would have naively guessed   based on the similar formula for the case of diagonal matrices $X$ that we    derived in appendix
\nref{sec:bfss_pert}, and
was mentioned in \nref{PotAp}.

\section{Analyzing the motion of a folded string} 

\la{TipMotion}

In this appendix we consider the motion of a folded stretched string. This is just a one parameter family of solutions out of the whole 
space of possible string motions.

\begin{figure}[h!]
\centering
\includegraphics{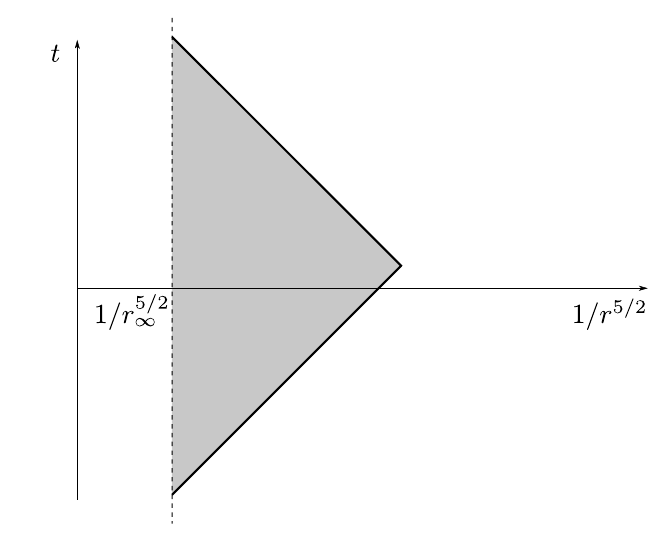}
\caption{Radial excitations of the adjoint string. The string tip moves close to a null-geodesic (bold line).
$r_\infty \sim \lambda^{1/3}$ is the region where the high curvature region starts. We imagine that
when the string reaches that point it bounces back with some reflection factor. }
\label{string_tip}
\end{figure} 

We can view the motion of the tip from the high curvature region to the low curvature and back as a kind of scattering problem. See 
Figure \ref{string_tip}. 
So we can calculate the total phase shift accumulated during the process via a WKB approximation. 

%We now return to strong coupling. Around the thermal background we have argued that the minimum 
%energy for the adjoint state is given in  \nref{Estimate}. We would now like to discuss excitations above this states. In the planar limit these excitations will be single strings of operators with a fundamental index at one end and an antifundamental at the other end, combinded so that we 
%adjoint index all together.  We expect that these states would be strings whose ends are located in the 
%high curvature region. 

%Of course, when we quantize the string we expect a large number of modes. For simplicity, here
%we consider a very particular case where the string motion looks like that some instant of time it looks 
%liked a folded string with a tip moving only along the radial direction. The full stpacetime motion is depicted in Figure   \ref{string_tip}. We can consider an approximate WKB quantization in the space of such motions. 
We view the system as the tip of a string which is approximated as a particle with large 
momentum $p$  and energy linear in the momentum. This tip is acted on by the rest of the string which provides a
potential. The full 
Hamiltonian is  
\be \la{HamTip}
H = \sqrt{-{g_{tt} \over g_{rr}} } |p|+\frac{r_\infty-r}{ \pi}
\ee
where $p$ is the momentum conjugate to $r$.  
The tip of the string starts from the large $r$ region with very high ingoing radial momentum, the string pulls and slows it down until the tip bounces back to the large $r$ region, see 
Figure \ref{string_tip}.
The total phase shift then is 
\be
 \delta_{\rm bulk } = 2 \int_{r_\infty-\pi E}^{r_\infty} dr \  p 
 % = 2 \pi n  - \delta , \qquad n \in \mathbb{Z}
\ee
Which can be re-written in terms of the total energy $E$ given in \nref{HamTip}.
This gives for $E \ra \cfrac{r_\infty}{ \pi}$:
\bea  
\delta_{\rm bulk } = \cfrac{8 \sqrt{\lm d_0} }{15 \pi (\pi E_\infty- \pi E_n)^{3/2}}
%&=& 2\pi (n - n_0)  ~~~~\Rightarrow 
%\cr
%E_n &\sim &% \cfrac{1}{\pi} \l r_\infty -  \l \cfrac{4 \sqrt{\lm d_0}}{ 15 \pi^2 n} \r^{2/3} \r = 
% \lm^{1/3} \left[ \widetilde{C} - { 1 \over \pi } \l \cfrac{4 \sqrt{ d_0}}{ 15 \pi^2( n-n_0)} \r^{2/3} \right] \la{ExSp}
\eea
Where $E_\infty \equiv  \tilde C \lambda^{1/3}$
 is the energy of a folded string that stretches all the way to $r=0$. For this reason we expect that $\tilde C > C$ by an order one amount. 
 A similar problem in the linear dilaton background that is dual to the double scaling limit of a single matrix model was analyzed in 
 \cite{Maldacena:2005hi} and matched to the matrix model computation in \cite{Fidkowski:2005ck}. 
 
 In order to figure out the whole motion, we need to know how the tip bounces back from the high curvature region. This seems to be a difficult problem since the state that comes out could have more excitations on the string worldvolume. Solving this would involve connecting the motion in the weakly curved region to the motion in the perturbative matrix model region. We will not do this here.
 Instead we will simply assume that the problem is such that the string tip comes back out with an extra phase shift $\delta_{\rm high}$ from the high curvature region. Furthermore we will assume that it is basically a constant for $E \sim E_{\infty}$. This is a non-trivial assumption and it is likely wrong. The only reason we make it is to define a toy problem where we can now semiclassically quantize the motion by setting 
 \be
 \delta_{\rm bulk} (E) + \delta_{\rm high} = 2 \pi n 
 \ee
 Leading to 
 \be
 E_n \sim   
 \lm^{1/3} \left[ \widetilde{C} - { 1 \over \pi } \l \cfrac{4 \sqrt{ d_0}}{ 15 \pi^2( n-n_0)} \r^{2/3} \right] \la{ExSp} ~,~~~~~~~ 
 n_0 \equiv { \delta_{\rm high}  \over 2 \pi }
 \ee
 Note that $n_0$ is not an integer.

We see that there is an infinite tower of excitations. 
 For non-zero temperature, there is actually an $n_{max} -n_0 \propto  T^{-3/5}$ where the states change behavior qualitatively, the folded string falls into the black hole horizon and
 stays there forever.  (At finite $N$ the string can break and the fold can return to infinity). 
 To describe this behavior we need to study the non-extermal metric (\ref{geometry}).
Now we have a non-extremal black hole with a horizon at $r=r_0$. If the 
string has enough energy to reach the horizon, then the tip will
fall into the black hole and never come back. This sets an upper bound for the
energy:
\be
E_\text{\rm dec} - E_{\infty} = - \cfrac{r_0}{\pi } 
\ee 
Moreover now we have a finite number of excited states that do not fall into the black hole
\be
 n_\text{max} -n_0  =\sqrt{\lm d_0} \int_{r_0}^{r_\infty} dr\ 
\cfrac{r-r_0}{\pi^2 r^{7/2}(1-r_0^7/r^7)} = 
\cfrac{4.06 \sqrt{\lm d_0}}{14 \pi^2 r_0^{3/2}} \sim T^{-3/5}, \ \text{for} \  r_\infty \gg r_0
\ee
 %
%Nonetheless, for low enough temperatures $T \ll \lm^{1/3}$ we do not see these higher excitations. 

For states with $E> E_{\rm dec }$, the string tip falls into the black hole and the state becomes a string and an anti-string,
 both ending at the horizon as independent excitations. 
 
\section{Scaling properties of the solution and the action}

\label{app:scaling}
In this Appendix we briefly discuss some scaling properties of the solution \nref{geometry}. We find that under the following 
rescaling of the coordinates the metric and the dilaton rescale as
% metric (\ref{geometry}). It is easy to see that the metric and the dilaton has the following symmetry:
\beq
%\begin{split}
%r \ra r \ \eta \\
%ds \ra ds \ \eta^{-3/4} \\
%t \ra t \ \eta^{-5/2} \\
%e^{2 \phi} \ra e^{2\phi } \ \eta^{-21/2} 
%\end{split} 
\begin{split}
t \to \eta  t ~,~~~~~~~ r \to  \eta^{ - 2/5} r \\
ds^2 \ra ds^2  \ \eta^{-3/5} ~,~~~~~~ e^{2 \phi} \ra e^{2\phi } \ \eta^{-21/5} \\ \la{rescS}
\end{split} 
\eeq
% Notice that this symmetry is present in the extremal case $r_0=0$ as well.
The gravity action scales as
\beq
\label{gr_action}
S_\text{gravity} = \int d^{10} x \ e^{-2 \phi} \sqrt{g} R \sim \eta^{-9/5} \sim T^{9/5}
\eeq
Notice that  $\beta $ is recaled when we rescale time. 
This is the correct behaviour of the Bekenstein--Hawking entropy (\ref{BHentropy}). Notice that the action and entropy scale in the same way. 
Notice that since the action changes \nref{rescS} is not a symmetry of the action, but it helps determine  the temperature dependence. 

It turns out that the  Dirac--Born--Infeld(DBI) action for a probe D0 brane in the \textit{extremal} geometry (\ref{geometry})
with $r_0=0$ has exactly the same scaling behaviour. This can be checked explicitly, but we can also derive it by the following observations.
%
%Let us explain why the DBI action for a D0 probe:
The action is 
\beq
\label{probe_dbi}
S_\text{DBI} = -\int e^{-\phi} ds + \int A_t dt
\eeq
%has exactly the same scaling as the gravity action (\ref{gr_action}). There are several steps:
We now observe:
\begin{itemize}
\item The derivative of the free energy with respect to the charge yields the difference between the RR 1-form at the horizon
and infinity\footnote{More precisely, one has to substract the zero-temperature value in order to make this expression finite.}:
\beq
\frac{\pr F}{\pr N} = A_t \at_\text{horizon} - A_t \at_\text{infinity}
\eeq
this is why $A_t$ scales as the free energy.
\item Notice that the expression for $A_t$ in (\ref{geometry}) does not 
contain $r_0$. This is why it has exactly the same scaling for both
extremal and non-extremal cases.
\item Finally, both terms in (\ref{probe_dbi}) scale in the same way as in the extremal case because of the supersymmetry
(there should be no force acting on a D0 brane at rest).
\end{itemize}

Now, this observation also explains why the following action has the same rescaling properties 
\be \la{actsi}
S = \int dt  \left[  \vec v _i^2 +  ({\rm const} ) { (\vec v_i - \vec v_j)^4  \over |\vec r_i - r_j|^7 }  \right] 
\ee
under \nref{rescS}. The reason is that the velocity expansion of \nref{probe_dbi} gives rise to a particular case of this action. 

The point of these observations is to ``explain'' the observation in 
%This proves that the D0 brane probe action (\ref{probe_dbi}) scales as the gravity action (\ref{gr_action}).
%This explains the observation made in 
\cite{Smilga:2008bt,soup} that \nref{actsi} has the same scaling as the entropy. The arguments used in that paper were scaling arguments, and
they have reproduced the entropy for simple scaling reasons. But it seems that the thermodynamics of \nref{actsi} is really ill defined because it has a
``fall to the center'' instability.

\mciteSetMidEndSepPunct{}{\ifmciteBstWouldAddEndPunct.\else\fi}{\relax}
\bibliographystyle{utphys}
\bibliography{UnGaugedMatrixModel}{}

\end{document}